\documentclass[a4paper,fleqn,usenatbib, useAMS]{mnras}

\usepackage[varg]{txfonts}
\usepackage[T1]{fontenc}
\usepackage{ae,aecompl}

\usepackage{graphicx}
\usepackage{footnote}
\usepackage{color, soul}

\usepackage[all]{hypcap}
\usepackage{caption}
\usepackage{sidecap}
\sidecaptionvpos{figure}{t}

\defcitealias{Khanzadyan2011}{K11}

\title[Connecting low- and high-mass star formation]{Connecting low- and high-mass star formation: the intermediate-mass protostar IRAS 05373+2349 VLA 2}

\author[G. M. Brown et al.]{
G. M. Brown,$^{1}$\thanks{E-mail: py11gb@leeds.ac.uk}
K. G. Johnston,$^{1}$
M. G. Hoare$^{1}$
and S. L. Lumsden$^{1}$
\\
$^{1}$School of Physics and Astronomy, University of Leeds, West Yorkshire, Leeds LS2 9JT, UK
}

\pubyear{2016}
   
\begin{document}
\label{firstpage}
\pagerange{\pageref{firstpage}--\pageref{lastpage}}
\maketitle

% Abstract of the paper
\begin{abstract}
Until recently, there have been few studies of the protostellar evolution of intermediate-mass (IM) stars, which may bridge the low-and high-mass regimes. This paper aims to investigate whether the properties of an IM protostar within the IRAS~05373+2349 embedded cluster are similar to that of low- and/or high-mass protostars. We carried out Very Large Array as well as Combined Array for Research in Millimeter Astronomy continuum and $^{12}$CO(J=1--0) observations, which uncover seven radio continuum sources (VLA~1--7). The spectral index of VLA~2, associated with the IM protostar is consistent with an ionised stellar wind or jet. The source VLA~3 is coincident with previously observed H$_{2}$ emission line objects aligned in the north-south direction (P.A. -20 to -12$\degr$), which may be either an ionised jet emanating from VLA~2 or (shock-)ionised cavity walls in the large-scale outflow from VLA~2. The position angle between VLA~2 and 3 is slightly misaligned with the large-scale outflow we map at $\sim$5-arcsec resolution in $^{12}$CO (P.A. $\sim$30$\degr$), which in the case of a jet suggests precession. The emission from the mm core associated with VLA~2 is also detected; we estimate its mass to be 12--23~M$_{\sun}$, depending on the contribution from ionised gas. Furthermore, the large-scale outflow has properties intermediate between outflows from low- and high-mass young stars. Therefore, we conclude that the IM protostar within IRAS~05373+2349 is phenomenologically as well as quantitatively intermediate between the low- and high-mass domains.
\end{abstract}

% Select between one and six entries from the list of approved keywords.
% Don't make up new ones.
\begin{keywords}
stars: formation -- stars: protostars -- ISM: jets and outflows -- radio continuum: stars -- techniques: interferometric
\end{keywords}

%%%%%%%%%%%%%%%%%%%%%%%%%%%%%%%%%%%%%%%%%%%%%%%%%%

%%%%%%%%%%%%%%%%% BODY OF PAPER %%%%%%%%%%%%%%%%%%

\section{Introduction} \label{Sec:Intro}

As much research has focused on either distinguishing or unifying the formation of high-mass stars and their low-mass counterparts, intermediate-mass (IM) stars are of interest as they provide a bridge between the two mass regimes. IM protostars -- the precursors of Herbig Ae/Be stars -- are defined as young stellar objects (YSOs) that will reach final masses of 2--8~M$_{\sun}$ and have luminosities between $\sim$50--2000~L$_{\sun}$ \citep{Beltran2015}. The lower limit of 2~M$_{\sun}$ originates from the fact that above this limit stars should not have the outer convective zones required to produce the magnetic fields needed for magnetically-mediated accretion \citep[see][for an observational confirmation of this temperature and thus mass limit]{Simon02}; the upper limit of 8~M$_{\sun}$ originates from the stellar mass required to produce a type II supernova, which is used to define the lower mass limit for high-mass stars \citep{Zinnecker07}. The upper mass limit also corresponds to the mass above which photo-ionization by UV photons becomes easily observable in cm continuum and the mass above which a pre-main sequence phase is not observable \citep{Beltran16}.

Unlike high-mass stars, whose descent onto the main sequence occurs while still deeply embedded and actively accreting within their parent cores, IM stars have a longer pre-main sequence timescale, so that their discs are revealed for part of their formation, similar to their low-mass counterparts. This is a result of their accretion timescales being shorter than their Kelvin-Helmholtz timescales, so that these stars are finished accreting before they reach the main sequence \citep{Beuther07}. On the other hand, IM stars form in more densely clustered environments, similar to high-mass stars \citep[e.g.][]{Fuente07,Gutermuth05}, with a smooth transition to the regime of rich clusters at a mass of 6~M$_{\sun}$ (\citealt{Testi99_new}). Thus, IM protostars should also share some of the characteristics of those at higher masses.

To date, there have been only a handful of high-resolution studies of IM protostars. One of the first, \citet{Fuente01} presented a study including two IM protostars: NGC~7129 FIR1 and FIR2, uncovering multiplicity and/or associated clusters, protostellar envelope masses of 2-3.5~M$_{\sun}$ and energetic bipolar outflows which appear to be driven by several sources. \citet{Beltran06} studied the morphology of the outflow from the IM protostar IRAS~22272+6358A, finding that  the 14.2~M$_{\sun}$ core OVRO~2, one of the four detected continuum sources in the region, was powering the asymmetric and collimated molecular outflow. The properties of the outflow were consistent with those of the outflows driven by low-mass YSOs. In a companion paper to \citet{Beltran02} and \citet{Beltran06} that studied IRAS~21391+5802 and IRAS~22272+6358A mentioned above, \citet{Beltran08} studied the 280~L$_{\sun}$ protostar IRAS~20050+2720. They detected three dust cores in the region at 2.7~mm (OVRO~1-3), and two bipolar flows in $^{12}$CO J = 1 -- 0, one of which is powered by OVRO~1. Combining their study of the outflows from these three IM protostars with a small number of objects from the literature, they concluded that although outflows from IM protostars are more energetic than those from low-mass protostars, they were not necessarily more complex, being collimated even at low velocities. In addition, they suggest that the increased outflow momentum rates compared to outflows from low-mass protostars are likely due to higher accretion rates for these objects. A further finding was that the IM protostars studied all formed in protoclusters containing several IM and low-mass millimetre sources.

More recently, \citet{van-Kempen16} used APEX to study a sample of six IM protostars, mapping the bipolar outflows and quiescent gas in $^{12}$CO and $^{13}$CO J=6 -- 5 and finding that their line luminosities and outflow forces also follow trends with bolometric luminosity and outflow mass, connecting low- and high-mass protostars, however they could not confirm the result of \citet{Beltran08} that fragmentation enhances outflow forces for IM protostars in clusters. Studying one of the largest samples thus far, \citet{Crimier10} determined the physical structure of the envelopes of a sample of five IM protostars via radiative-transfer modelling, finding that the physical parameters describing the envelopes vary smoothly between low- and high-mass protostars, and that the density structure was consistent with predictions from the predictions of ``inside-out'' collapse.

Other IM protostars studied to date include IRAS 22198+6336 \citep{Sanchez-Monge10, Palau11}, AFGL 5142 \citep{Palau11}, IC 1396 N \citep{Neri07, Fuente09}, 13 S in R CrA \citep{Saul15}, MMS 6/OMC-3 \citep{Takahashi12a, Takahashi12b}, IRAS 05345+3157 \citep{Fontani09} and G173.58+2.45 \citep{Shepherd02}. In this work, our aim is to increase this relatively small number by investigating the physical properties of the IM protostar associated with IRAS~05373+2349, which has a luminosity of 290-600 L$_{\odot}$ \citep{Khanzadyan2011, Molinari2000, Zhang2005}, to uncover the differences and similarities to low- and high-mass star formation. To do this, we have carried out the first mm-interferometric observations of IRAS~05373+2349, which lies at a distance of $\sim$1.2~kpc \citep{Molinari2000}. The associated embedded cluster and the nearby GGD 4 object have been the focus of several studies due to the active star formation occurring within the region. \citet{Zhang2005} has mapped the large-scale outflow from IRAS~05373+2349 in $^{12}$CO\,(J=1--0) with 29-arcsec resolution, determining that it has a NE-SW orientation. A more recent study by \citet[][hereafter \citetalias{Khanzadyan2011}]{Khanzadyan2011} identified numerous possible outflows in the immediate area of IRAS~05373+2349 based on near-infrared H$_2$ line emission and investigated the candidate driving sources. 
 
In this paper, we present continuum observations of IRAS 05373+2349 at 6, 3.6, 1.3-cm and 2.7-mm wavelengths as well as $^{12}$CO\,(J=1--0) line emission. In \autoref{Sec:Obs} we describe the observations. In \autoref{Sec:Results} we present our results, including the derived outflow properties. We present our discussion in \autoref{Sec:Discussion} and in \autoref{Sec:Conclusion} we outline our conclusions.

%__________________________________________________ Two column table
   \begin{table*}  % substistute table with table* for 2 column wide
      \caption{Summary of Continuum Observations.}
         \label{Tab:Obs Summary}
     	 \center
         \begin{tabular}{lllllllll}
            \hline
            \noalign{\smallskip}
            Wavelength & Array & Date of   & No. of &  Time on-&  Synthesised & P.A & Map rms \\
                       &   & observation  & antennas & source (hr) & beam size (") & ($^{\circ}$) & (mJy beam$^{-1}$)\\
            \noalign{\smallskip}
            \hline
            \noalign{\smallskip}
            6~cm   & VLA-C & 12 Mar 2008   & 27 (24) & 1.5 &  4.55 $\times$ 4.02 & 27 & 0.026 \\
            3.6~cm & VLA-C & 10 Mar 2008 & 26 (23) & 2.0 &  2.97 $\times$ 2.86 & -13 & 0.016\\
            1.3~cm & VLA-C & 10 Mar 2008 & 26 (21) & 1.3 &  1.23 $\times$ 1.04 & 48 & 0.047\\
            2.7 mm & CARMA-D & 24 Apr 2007 & 15      & 5.7 &  4.92 $\times$ 4.28 & 78 & 2.6 \\
         
            \noalign{\smallskip}
            \hline
         \end{tabular}
     
   \end{table*}
%________________________________________________________________ 

%____________CALIBRATOR TABLE______________________________________________________

   %__________________________________________________ One column table
\begin{table}  
\caption{Assumed and determined flux and phase calibrator properties.}
\label{Tab:Cals}
\center    	
\begin{tabular}{lccccc} 
\hline
\noalign{\smallskip}
Source & \multicolumn{4}{c}{Flux density (Jy)} &  Cal  \\
	   & 6~cm & 3.6~cm  & 1.3~cm & 2.7 mm & type  \\
\noalign{\smallskip}
\hline
\noalign{\smallskip}
0137+331 & 5.52 & - & - & - & A  \\
1331+305 & - & 5.22 & 2.58 & - & A  \\
0530+135 & - & - & - & 6.70 & A+G \\
0559+238 & 0.41 & 0.37 & - & - & G  \\
0539+145 & - & - & 0.35 & - & G  \\		
\noalign{\smallskip}
\hline
\end{tabular}    
\begin{description}
\item Absolute flux calibrators are denoted by `A' and the gain calibrators are  denoted by `G' in the `cal type' column.
\end{description}
\end{table}

\section{Interferometric Observations} \label{Sec:Obs}

\subsection{VLA 6, 3.6 and 1.3-cm continuum}

Multi-wavelength radio continuum observations were taken of IRAS 05373+2349 over two days during March 2008, with the Very Large Array (VLA) of the National Radio Astronomy Observatory\footnote{The National Radio Astronomy Observatory is a facility of the National Science Foundation operated under cooperative agreement by Associated Universities, Inc.}. The program number for these observations was AJ337. The pointing centre was 05$^{\rmn{h}}$40$^{\rmn{m}}$24$\fs$40 +23$\degr$50$\arcmin$54$\farcs$00 (J2000).  

The first set of observations in the X($\lambda$ = 3.6-cm) and K($\lambda =$ 1.3-cm) bands were taken on 10 March 2008 and the observation of the C($\lambda$ = 6-cm) band was taken on 12 March 2008. For each observation two 50-MHz spectral windows were placed at: 4.89 and 4.84 GHz, 8.44 and 8.49 GHz, and 22.5 and 22.4 GHz for the 6, 3.6 and 1.3-cm bands, respectively. The observations were performed in the VLA's C configuration which had baseline lengths between 35~m and 3.2~km, which produced information on angular scales from 3.88, 2.31 and 1.4~arcsec to 4.9, 3.0 and 1.4~arcmin for 6, 3.6 and 1.3~cm, respectively. 

\autoref{Tab:Obs Summary} presents a summary of the observations which lists the observed wavelength, configuration, observation date, number of antennas (with the number of antennas with useful data given in parentheses), time on-source, synthesised beam size, position angle (P.A.) and the map rms noise.

\autoref{Tab:Cals} presents the calibrators used in the observations; the `Cal type' column showing their application. \autoref{Tab:Cals} also presents the fluxes derived from the available models for the flux calibrators 3C48 (0137+331) and 3C286 (1331+305), and the bootstrapped fluxes derived from the gain calibrators. 

Data reduction and imaging was carried out using the Common Astronomy Software Applications (\textsc{casa}) package version 4.2.2 \citep{CASAREF}. Briggs weighting with a robust parameter of 1.5 was used to clean the images, to retain optimal sensitivity in the observations. Data from baselines shorter than 50~m were removed from the 6-cm image to eliminate flux that was partially resolved-out. 

\subsection{CARMA $^{12}$CO\,(J=1--0) and 2.7-mm continuum}

Combined Array for Research in Millimeter Astronomy (CARMA\footnote{Support for CARMA construction was derived from the states of California, Illinois, and Maryland, the James S. McDonnell Foundation, the Gordon and Betty Moore Foundation, the Kenneth T. and Eileen L. Norris Foundation, the University of Chicago, the Associates of the California Institute of Technology, and the National Science Foundation. CARMA development and operations were supported by the National Science Foundation under a cooperative agreement, and by the CARMA partner universities.}) observations at $\lambda$ = 2.7~mm were taken on 24 April 2007. The antenna array during the observation consisted of fifteen antennas, six 10~m and nine 6~m in diameter with primary beams sizes of 64~arcsec and 115~arcsec, respectively. A mosaic was created from nine pointing centres in a square grid pattern. The pointings were separated by 30~arcsec.  

The CARMA correlator was set up with two sidebands, placed either side of the chosen local oscillator frequency of 113.280~GHz, with the upper bands situated to measure $^{12}$CO\,(J=1--0) line emission. Both the upper and lower sidebands contained one wide and two narrow spectral windows, giving a total of six windows. The wide spectral windows each had bandwidths of 468.750~MHz, a total of 15~channels with a spectral resolution of 31.250~MHz and were centred on 111.558 and 115.000~GHz, respectively. The two narrow windows in each sideband had bandwidths of 30.76 and 7.69~MHz, 63~channels and spectral resolutions of 488.281 and 122.070~kHz. The narrow windows were centred on 111.288~GHz in the lower sideband and 115.271~GHz in the upper sideband.

Hanning smoothing and line length corrections were applied to the data in \textsc{miriad}, which gave spectral resolutions of 1.27 and 0.32~km\,s$^{-1}$ for the two narrow windows covering the $^{12}$CO\,(J=1--0) line emission. The data were then exported from their original format into \textsc{casa} where data reduction and imaging were carried out. The source 0530+135 was used to calibrate both amplitude and gain of the observation; the assumed flux is listed in \autoref{Tab:Cals}. 

The images created from the millimetre continuum and $^{12}$CO\,(J=1--0) line emission are presented in \autoref{Subsec: CARMA results}. The two continuum spectral windows were cleaned together to produce a single image with a central frequency of 113.279~GHz, using Briggs weighting with a robust parameter of 0.5. The synthesised beam size and noise, along with the number of antennas and time spent on-source for the observation, can be found in \autoref{Tab:Obs Summary}. 

For the $^{12}$CO\,(J=1--0) line image, the two narrow spectral windows were combined at 115.271~GHz using natural weighting and a channel width of 1.3~km\,s$^{-1}$; uv--distances smaller than 3~k$\lambda$ were excluded. The beam size of the $^{12}$CO\,(J=1--0) image was 4.92\,$\times$\,4.28~arcsec P.A. 78$^{\circ}$  and the noise ranged from $\sim$0.07~Jy\,beam$^{-1}$ in an empty channel, up to $\sim$0.5~Jy\,beam$^{-1}$ in the inner channels.

\begin{table*} 
\caption{Measured properties of the observed 6-cm (5GHz), 3.6-cm (8.5GHz) and 1.3-cm (22.5GHz) continuum sources.}
\center     	
\begin{tabular}{lccccc c cc} 
\hline
\noalign{\smallskip}
Source & $\lambda$  & \multicolumn{2}{c}{Peak position} & Peak flux density  & Integrated flux & Deconvolved  & P.A. & Convolved \\
name & (cm) & RA (J2000) & Dec. (J2000) & (mJy beam$^{-1}$) & density (mJy) &  size ($\arcsec$) & ($^{\circ}$)  & size ($\arcsec$)\\
\noalign{\smallskip}
\hline
\noalign{\smallskip}
VLA 1 & 6   & 05:40:21.31 & 23:52:09.80 & 0.22 $\pm$ 0.03 & 0.40 $\pm$ 0.06 & 5.6 $\times$ 2.0 & \,\,\,84 $\pm$ 173  & 9.2 \\
      & 3.6 & 05:40:21.28 & 23:52:10.00 & 0.19 $\pm$ 0.02 & 0.29 $\pm$ 0.03 & 2.5 $\times$ 1.6 & 49 $\pm$ 20   & 5.6 \\
      & 1.3 & ...         & ...         \\
VLA 2 & 6   & 05:40:24.22 & 23:50:54.80 & 0.42 $\pm$ 0.02 & 0.45 $\pm$ 0.04 & 1.6 $\times$ 0.4 & \,\,\,39 $\pm$ 124 & 8.5 \\
      & 3.6* & 05:40:24.23 & 23:50:54.83 & 0.52 $\pm$ 0.03 & 0.85 $\pm$ 0.08 & - & -  & 7.1 \\
      & 1.3* & 05:40:24.22 & 23:50:54.76 & 0.46 $\pm$ 0.03 & 0.57 $\pm$ 0.12 & - & - & 2.5  \\
VLA 3 & 6   & 05:40:24.16 & 23:51:01.41 & 0.13 $\pm$ 0.02 & 0.15 $\pm$ 0.02 & <\,2.1  & - & 4.6  \\
      & 3.6 & 05:40:24.10 & 23:51:02.51 & 0.11 $\pm$ 0.02 & 0.19 $\pm$ 0.03 & 2.9 $\times$ 1.9 & 15 $\pm$ 19 & 5.3 \\
      & 1.3 & ...         & ...         \\
VLA 4 & 6   & 05:40:08.48 & 23:51:18.01 & 0.30 $\pm$ 0.04 & 0.72 $\pm$ 0.13 & <\,1.3  & - & 14.2  \\
      & 3.6 & ...         & ...         \\
      & 1.3 & ...         & ...         \\
VLA 5 & 6   & 05:40:10.52 & 23:49:32.75 & 0.93 $\pm$ 0.03 & 0.93 $\pm$ 0.07 & <\,0.8 & -  & 8.6  \\
      & 3.6 & 05:40:10.49 & 23:49:32.57 & 0.57 $\pm$ 0.03 & 0.76 $\pm$ 0.07 & 1.8 $\times$ 1.6 & 101 $\pm$ 168 & 4.5 \\
      & 1.3 & ...         & ...          \\
VLA 6 & 6   & 05:40:13.74 & 23:49:10.68 & 0.34 $\pm$ 0.02 & 0.34 $\pm$ 0.02 & <\,1.3 & -  & 6.3  \\
      & 3.6 & 05:40:13.74 & 23:49:09.54 & 0.26 $\pm$ 0.02 & 0.32 $\pm$ 0.03 & 1.9 $\times$ 0.6 & 24 $\pm$ 37 & 3.5 \\
      & 1.3 & ...         & ...         \\
VLA 7 & 6   & 05:40:08.02 & 23:47:22.78 & 2.06 $\pm$ 0.06 & 2.28 $\pm$ 0.18 & <\,0.5 & - & 10.4  \\
      & 3.6 & 05:40:07.97 & 23:47:22.22 & -               & -               &  -    & -             \\
      & 1.3 & ...         & ...         \\
\noalign{\smallskip}
\hline
\end{tabular}
\begin{description}
\item * indicates sources that were measured in an aperture defined by the 1$\sigma$ contour, as they were not well-fitted by a Gaussian. 
\item An ellipsis (...) in the peak position columns indicate non-detections at that wavelength. 
\item A dash indicates the property was unable to be determined. The 3.6~cm flux of VLA~7 could not be accurately determined as it was not possible to apply a correction for the primary beam response at this position.  
\end{description}
\label{Tab:Prop Full}  	
\end{table*}

\section{Results} \label{Sec:Results}

\subsection{VLA 6, 3. and 1.3-cm continuum} \label{Subsec:Results intro}

In this section, we present the VLA images at each wavelength and calculate spectral indices. 

\begin{figure*}
   \centering
   \includegraphics[width=\hsize]{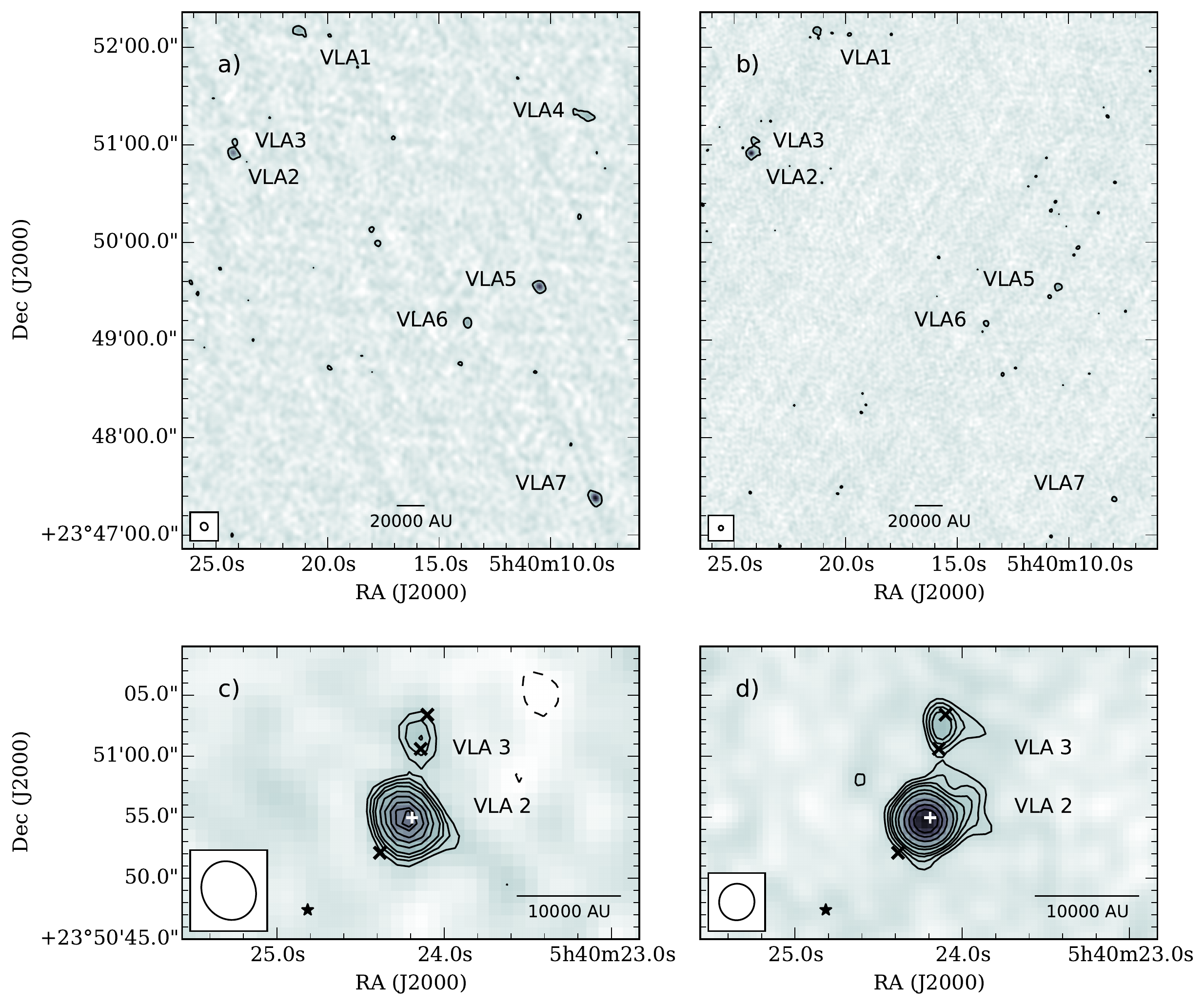}
   \caption{Continuum emission at 6~cm (left column, $\sigma$ = 26~$\umu$Jy\,beam$^{-1}$) and 3.6~cm (right column, $\sigma$ = 16~$\umu$Jy\,beam$^{-1}$). Panels a and b show the region surrounding the IRAS source. Panels c and d show VLA~2 and 3. The position of the millimetre core seen at 2.7~mm (see \autoref{Fig:mm continuum}) associated with VLA~2 is marked by a plus. A Class~I object reported by \citep{Gutermuth2009} is marked by a black star. The black x's indicate the positions of MHO 745 A, B and C detected by \citet{Khanzadyan2011}.  
   Contour levels show for (a and b): -3, 3 $\times$ $\sigma$, (c): -3, 3, 4, 5, 6, 8, 10, 12, 15 $\times$ $\sigma$, (d): -3, 3, 4, 5, 6, 8, 10, 12, 15, 20, 25 $\times$ $\sigma$. The beam and scale bar are shown at the bottom of the images.}
         \label{Fig:3.6 and 6 cm} 
\end{figure*}

The radio continuum maps are presented in \autoref{Fig:3.6 and 6 cm}, which shows the 6 and 3.6-cm continuum emission with a wide view of the surrounding region (panels a and b), and towards VLA~2 and 3 (panels c and d), and in \autoref{Fig:1.3cm}, which shows the 1.3-cm continuum towards VLA~2.

\autoref{Tab:Prop Full} presents the peak position, integrated and peak flux densities, the deconvolved source size and position angle and the convolved size measured as the largest extent of the 3$\sigma$ contours. The properties presented in \autoref{Tab:Prop Full} were derived from fitting a 2D Gaussian to each radio source, with the exception of two non-Gaussian sources which are indicated by an * in the $\lambda$ column. 

For Gaussian sources unresolved by the fit, an upper size limit is given in the deconvolved source size column of \autoref{Tab:Prop Full}. The upper size limit is determined by calculating the maximum possible source size when unresolved, which depends on the signal-to-noise of the source and the synthesised beam size of the observation.

For non-Gaussian sources, the peak flux density and position were measured at the brightest pixel and the integrated flux density was measured by summing up the emission within a region defined by the 1$\sigma$ contour of the source. 

%                                                One column figure
%----------------------------------------------------------- S_vib

The fitted source positions had an average error of $\sim$0.1$\arcsec$. The errors in the integrated flux of the non-Gaussian sources were determined by measuring the noise in nearby empty regions of sky using the same aperture shape used for the flux measurements.

The spectral index, $\alpha$, defined as
\begin{equation}
	\centering
	S_\nu \propto \nu^\alpha.
	\label{Eqn: spx ind}
\end{equation}
was determined for all sources where there were fluxes at more than one wavelength. In the case of VLA~2, which had three fluxes, the spectral index was derived from a least squares fit in log space, including errors in the fit. The spectral indices are given in Sections~\ref{Subsec:Results VLA 2 and 3} and \ref{subsec: Other results} below. The angular size--frequency index, $\xi$, given by
\begin{equation}
	\theta_\nu \propto \nu^\xi.
	\label{Eqn: size ind}
\end{equation}

The two indices ($\alpha$ and $\xi$) provide further classification of the emission coming from a thermal or non-thermal source. For thermal sources with an $\alpha$ = 0.6, which is related to the power-law relationship between electron density and emitting object radius, it can be shown that $\xi$ = -0.7 \citep{Panagia1975}. Whereas, non-thermal synchrotron emission, detected using interferometric techniques, can produce $\xi$ $\leq$ -2 \citep{Thompson1986}. We determined $\xi$ for VLA~1, which was resolved at more than one wavelength, using a similar fitting method as described above.

\subsubsection{VLA 2 and VLA 3} \label{Subsec:Results VLA 2 and 3}

\begin{figure}
   \centering
   \includegraphics[width=\hsize]{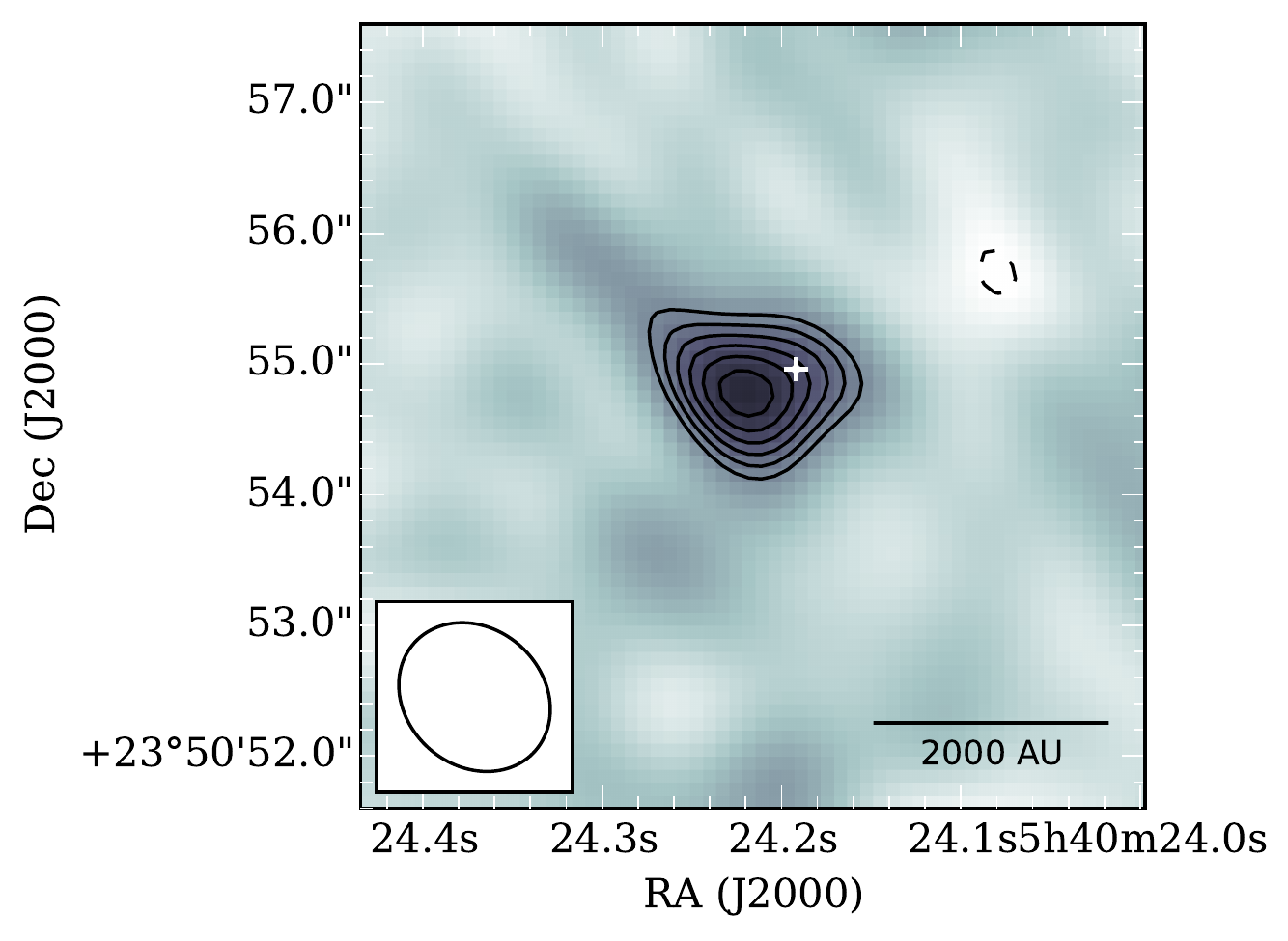}
      \caption{A close-up map of the 1.3-cm continuum emission towards VLA~2 with the position of the millimetre core seen at 2.7~mm (see \autoref{Fig:mm continuum}) associated with VLA~2 is marked by a plus. Contour levels are -3, 4, 5, 6, 7, 8 and 9 $\times$ 47 $\umu$Jy\,beam$^{-1}$. The beam is shown in the bottom left and a scale bar in the bottom right.}
         \label{Fig:1.3cm}
   \end{figure}

VLA~2 is associated with one of the most prominent members of the embedded cluster towards IRAS 05373+2349 and was detected at all wavelengths. \autoref{Fig:3.6 and 6 cm}\,(c and d) shows the 6 and 3.6-cm emission, respectively. VLA~2 was the only source detected at 1.3~cm, which is shown in \autoref{Fig:1.3cm}. 

The markers in these figures show: the peak position of the millimetre core associated with VLA~2 and 3 detected with CARMA (see \autoref{Subsec: CARMA results}), marked as a plus sign; a Class I protostar \citep[to the south--east of VLA~2,][]{Gutermuth2009}, marked as a black star, and the positions of three molecular hydrogen emission line objects \citepalias[MHO 745A, B and C,][]{Khanzadyan2011}, marked as crosses. 

The spectral index of VLA~2 was $\alpha$\,=\,0.38 $\pm$ 0.14, measured from fitting a line to the three fluxes\footnote{The integrated and peak flux densities of VLA~2 at 6~cm, measured in a 1$\sigma$ region, was 0.43$\pm$0.06~mJy and 0.41$\pm$0.03~mJy\,beam$^{-1}$, respectively. These are very similar to the fluxes derived from the Gaussian fit given in \autoref{Tab:Prop Full}, thus the determined spectral index is not affected by different methods of flux measurement.} given in \autoref{Tab:Prop Full}. As the 1.3-cm image, shown in \autoref{Fig:1.3cm}, is sensitive to more compact emission than 6 and 3.6-cm maps, while being less sensitive to the extended emission, it may be that we do not detect a significant portion of the flux of VLA~2 at this wavelength. Therefore, it is possible that the spectrum including the 1.3-cm flux measurement is artificially flattened. Ignoring the 1.3-cm flux yields a much larger spectral index of $\alpha$\,=\,1.20 $\pm$ 0.24. Both of these spectral indices are consistent with free-free emission of varying degrees of optical thickness.

VLA~2 was detected previously at 3.6~cm with a flux density of 0.70 ($\pm$ 0.04)~mJy and an estimated spectral index of 0.9 \citep{Molinari2002}. Our measured flux and spectral index for VLA~2 is consistent with this finding.

VLA~3 is a new radio source, 6.7~arcsec and 7.9~arcsec (at 6 and 3.6~cm, respectively) to the north of VLA~2. These correspond to separations of $\sim$8000 and $\sim$9500~au at 1.2~kpc. VLA~3 was resolved in our 3.6-cm observation, being extended roughly north-south with a P.A. of 15~deg. The spectral index for VLA~3 was found to be $\alpha$\,=\,0.45 $\pm$ 0.39 indicating partially optically thick free-free emission. We find no UKIDSS emission associated with this source, which would indicate a separate protostar.

\autoref{Fig:3.6 and 6 cm}\,(c and d) show the inner contours of VLA~3 coincide closely with the positions of MHO 745~A and B. \citetalias{Khanzadyan2011} suggested the MHOs (745 A, B and C) are powered by a bipolar jet originating from VLA~2 aligned in the north--south direction. We measured the position angle of this jet to be between -20 and -12$\degr$, where the former value was found by drawing a line between MHO~745~A and VLA~2 and the second was found between MHO~745~A and C (to the south-east of VLA~2).

The suggested jet is aligned within roughly 40\,--\,50$\degr$ of the large-scale outflow, which was found to have a position angle of $\sim$30$^{\circ}$ (see \autoref{subsubsec: 12CO results}). This is consistent with the classification criterion for an ionised jet used by \citep{Purser2016}, that the difference in deconvolved position angle between the jet and the outflow should be less than 45$^{\circ}$. Therefore the radio continuum emission from VLA~2 and VLA~3 can be explained as a single compact source plus a jet elongated to the north in both 6 and 3.6-cm maps. However, due to its misalignment with the large-scale outflow, another explanation could be that VLA~3 is tracing the (shock-)ionised cavity walls of the large-scale outflow (see \autoref{subsec:discussion small Jet}).

%----------------------------------------

\subsubsection{Other sources} \label{subsec: Other results}

The remaining detected sources are not known to be associated with IRAS~05373+2349. Following equation A2 of \citet{Anglada1998}, which gives the number of background sources as a function of integrated flux based on data from \citet{Condon1984}, we expect $\sim$2 background sources within the field of 4.5 by 5.5~arcmin at 6~cm, shown in \autoref{Fig:3.6 and 6 cm}\,(a).

VLA~1 is the first of two radio sources detected by \citet{Molinari2002}. There is no infrared source found to be immediately associated with VLA~1 in any \textit{Spitzer} IRAC \citep{IRAC} or UKIDSS band. The source has a spectral index of $\alpha$\,=\,-0.61\,$\pm$\,0.38 which indicates synchrotron emission. 

An angular size--frequency index, $\xi$\,=\,-0.96\,$\pm$\,0.92, was also derived for VLA~1. This index can be used to determine characteristics of the source. \citet{Yang2008} found that extragalactic jets show angular size--frequency indices of $\xi$\,=\,-0.95\,$\pm$\,0.37. The uncertainty in the measured angular size--frequency index is too large to draw such a conclusion. However, as there is no evidence of a YSO at this position, we find the source is likely extragalactic in origin. 

The detailed analysis of the H$_2$ line emission by \citetalias{Khanzadyan2011} did not focus as far afield as to where VLA~4--7 are located. Thus, it is difficult to associate any MHOs with these sources. 

VLA~4 was found to be clearly associated with IRAC emission at 3.6, 4.5, 5.8 and 7.9~$\micron$ and the \textit{K}-band of the UKIDSS survey. A spectral index could not be determined as it was only detected at 6~cm in our observations. However, a non-detection at 3.6~cm of VLA~4 indicates this source likely has a negative spectral index.

VLA~5 was found to be associated with faint IRAC emission at 3.6 and 4.5~$\micron$. A spectral index of $\alpha$\,=\,-0.38\,$\pm$\,0.24 was found, consistent with a non-thermal source or narrowly consistent with optically-thin free-free emission.

There is faint IRAC emission associated with VLA~6 at 3.6 and 4.5~$\micron$, which has a spectral index of $\alpha$\,=\,-0.15\,$\pm$\,0.28. It could equally be associated with a non-thermal or optically thin thermal source due to the uncertainty in the spectral index. 
 
VLA~7 was found to be associated with faint IRAC emission at 3.6, 4.5 and 5.8~$\micron$. The 3.6-cm flux of VLA~7 could not be accurately determined. This was due to the primary beam response not being accurately known so far from the pointing centre of the observation. However, it can be seen from our observations that the 3.6-cm detection of VLA~7 is much weaker than that at 6~cm, and therefore it likely has a negative spectral index.

Further observations of these sources will be required to determine their association with the IRAS~05373+2349 cluster or whether they are unrelated or extragalactic objects.  

%____2.7 mm continuum emission ______________________________
   \begin{figure}
   \centering
   \includegraphics[width=\hsize]{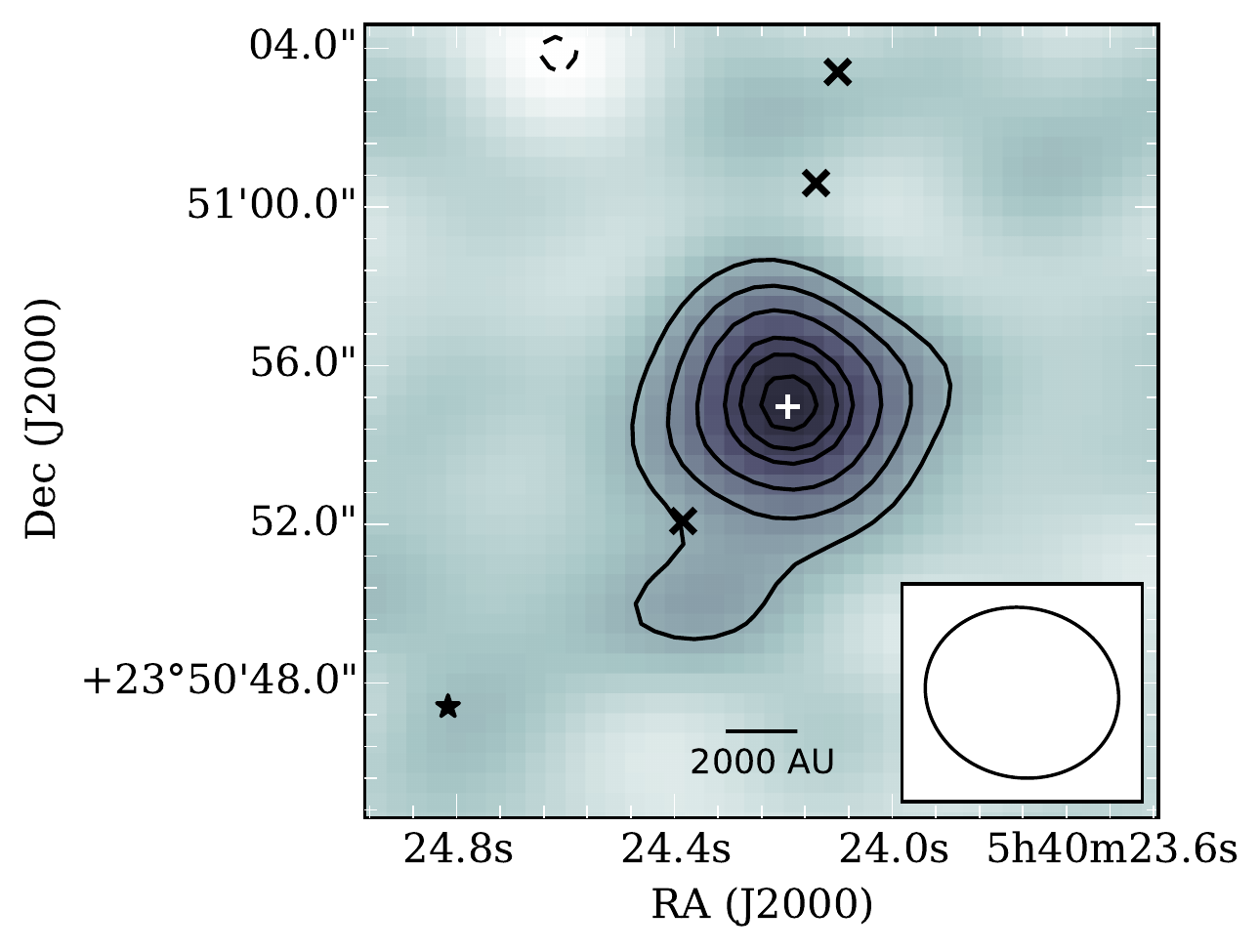}
      \caption{Continuum emission towards VLA~2 at 111.1GHz ($\sim$2.7 mm) shown greyscale and contours. The rms noise in the image is $\sigma$ = 2.6 mJy\,beam$^{-1}$. The greyscale extends from -3 $\times$ $\sigma$ to 27 mJy\,beam$^{-1}$. Contours are at -3, 3, 4, 5, 7 and 9 $\times$ $\sigma$. The synthesised beam is 4.9\,$\times$\,4.28~arcsec PA = 78$\degr$. Markers are the same as \autoref{Fig:3.6 and 6 cm}\,c.} 
         \label{Fig:mm continuum}
   \end{figure}
%
%______________________________________________________________

%____CO line spectra ____________________________
   \begin{figure}
   \centering
   \includegraphics[width=\hsize]{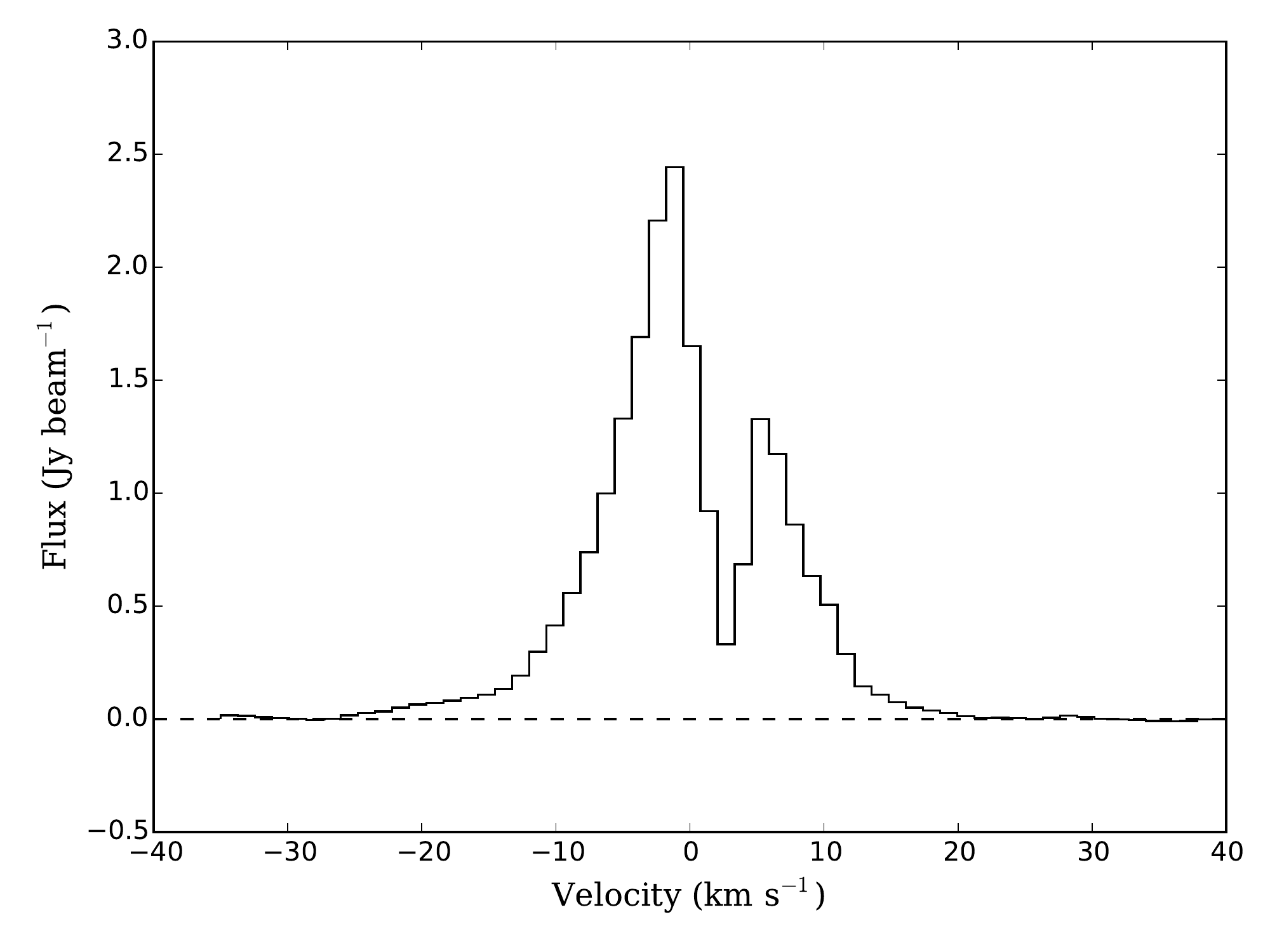}
      \caption{$^{12}$CO\,(J=1--0) spectral profile measured at the position of VLA~2: 05$^{\rmn{h}}$40$^{\rmn{m}}$24$\fs$22 +23$\degr$50$\arcmin$54$\farcs$79 (J2000).} 
         \label{Fig:CO spectrum}
   \end{figure}
%
%______________________________________________________________

\subsection{CARMA $^{12}$CO\,(J=1--0) and 2.7-mm continuum} \label{Subsec: CARMA results}

In this section the $^{12}$CO\,(J=1--0) line and millimetre continuum observations taken towards VLA~2 are used to infer the properties of the dense molecular gas surrounding the region. 

\subsubsection{2.7-mm continuum}

Continuum emission at 2.7~mm was only detected towards VLA~2. \autoref{Fig:mm continuum} shows the continuum emission from the region around VLA~2 at 111.1~GHz ($\sim$\,2.7~mm) with MHOs 745 A, B and C \citepalias{Khanzadyan2011} indicated as crosses. The morphology of the emission shows a compact core with an extension from the south of the source in the south--east direction. The integrated and peak flux densities, measured in a 1$\sigma$ contour, were 53.7\,$\pm$\,2.6~mJy and 30.1\,$\pm$\,1.6~mJy, respectively. The peak position, measured at the brightest pixel, was 05$^{\rmn{h}}$40$^{\rmn{m}}$24$\fs$18 +23$\degr$50$\arcmin$55$\farcs$01 (J2000). 

A 2D-Gaussian was also fitted to the source. However, we found that only the inner compact core could produce a satisfactory fit; we derived an integrated flux density of 53.7\,$\pm$\,5.9~mJy, a peak flux density of 29.9\,$\pm$\,3.3~mJy and a deconvolved source size of 4.1\,$\pm$\,0.4 by 2.9\,$\pm$\,0.6~arcsec and a P.A. of 162\,$\pm$\,18$\degr$. The deconvolved size corresponds to 4900\,$\times$\,3500~au at 1.2~kpc. The peak position was 05$^{\rmn{h}}$40$^{\rmn{m}}$24$\fs$194 +23$\degr$50$\arcmin$55$\farcs$346 (J2000), with a positional uncertainty on the order of 0.5$\arcsec$ \citep{Ikarashi2011}. Therefore the 2.7-mm source is associated with VLA~2.

To investigate how much millimetre emission might arise from the ionised material, we extrapolated the combined 6-cm flux of VLA~2 and VLA~3 to 2.7~mm, assuming a spectral index of $\alpha$\,=\,0.38\,$\pm$\,0.14 and that the spectrum did not turn over. We found an expected flux density of $\sim$1.95\,$\pm$\,0.73~mJy, which is $\sim$4 per cent of the measured millimetre continuum ($\sim$54~mJy). Therefore, the contribution of ionised gas emission in this case is negligible. However, if we use $\alpha$\,=\,1.20\,$\pm$\,0.24, determined by excluding the 1.3~cm fluxes, the estimated contribution to the combined integrated flux of VLA~2 and 3 at 2.7~mm becomes $\sim$25~mJy, roughly half of the measured millimetre continuum. Scaling the 6~cm fluxes of VLA~2 and 3 by $\alpha$\,=\,1.2 and $\alpha$\,=\,0.45, respectively, where the latter is the measured spectral index for VLA~3, gives a total 2.7~mm flux of $\sim$20~mJy. We calculated the core mass for both cases below. 

The mass of the material surrounding VLA~2, traced by the millimetre continuum emission was calculated using
\begin{equation}
\centering
M_{\mathrm{dust}} = \frac{d^{2}\,S_{\nu}}{B_{\nu}(T_{\mathrm{D}})\,\kappa_{\nu}}
\label{Eqn: Core mass}
\end{equation}
where $d$\,=\,1.2~kpc and $T_\mathrm{D}$ = 27~K are the distance and dust temperature \citep{Molinari2000}. $S_{\nu}$ is the integrated flux density measured in the 1$\sigma$ contour at $\lambda$\,=\,2.7~mm. $B_{\nu}$($T_{\mathrm{D}}$) is the value of the Planck function at $T_\mathrm{D}$. Finally the dust opacity, $\kappa_{\nu}$ = 0.1807~cm$^{-2}$\,g$^{-1}$, was found by extrapolating model opacity data \citep{Ossenkopf1994} to 2.7~mm, assuming a gas density of 1$\times$10$^{5}$~g\,cm$^{-3}$ and thin ice mantles, where $\beta$, the dust opacity index with frequency, was fitted to be -1.82. A gas-to-dust ratio of 100 was assumed to infer the total core mass. In the case of all of the 2.7~mm emission originating from the dust, the total core mass is 23~M$_{\sun}$. This is in close agreement with the value of $\sim$26~M$_{\sun}$ found by \citet{Molinari2008} from fitting the SED of the source. In the case of 25~mJy of the 2.7~mm emission originating from the ionised gas, and 29~mJy from the dust, we determined the core mass to be 12~M$_{\sun}$.

%____________OUTFLOW PROPERTIES TABLE______________________________________________

   %__________________________________________________ One column table
\begin{table} 
\caption{Derived properties of the blue- and red-shifted lobes of the outflow centred on VLA~2 traced by $^{12}$CO\,(J=1--0) emission.}
\center    	
\begin{tabular}{lcccc}
\hline
\noalign{\smallskip}
Parameter &  \multicolumn{2}{c}{Outflow} \\
	   &  blue lobe & red lobe \\
\noalign{\smallskip}
\hline
\noalign{\smallskip}
Velocity range (km\,s$^{-1}$) & -25.9 to -1.2 & 5.3 to 20.9 \\
\noalign{\smallskip}
Mass, $M$ (M$_{\sun}$) &  0.53 & 0.26 \\
Momentum, $P$ (M$_{\sun}$\,km\,s$^{-1}$) &  4.06 (4.82) & 1.57 (1.87) \\
Energy, $E$ (M$_{\sun}$\,km$^{2}$\,s$^{-2}$) & 20.70 (29.24) \, & 5.80 (8.19)\\
Energy, $E$ (10$^{44}$ erg) & 4.14 (5.85) & 1.16 (1.64) \\	
v$_{outflow}$ (km\,s$^{-1}$) & 7.61 (9.04) & 5.92 (7.03) \\
\noalign{\smallskip}
Length, $L$ (au) & 35000 (41000) & 43000 (51000) \\
\noalign{\smallskip}
t$_{dyn}$ (10$^{4}$ yr) & 2.15 (2.15) & 3.41 (3.41) \\
$\dot{M}$ (10$^{-5}$ M$_{\sun}$ yr$^{-1}$) & 2.48 (2.48) & 0.78 (0.78) \\
F (10$^{-4}$ M$_{\sun}$\,km\,s$^{-1}$ yr$^{-1}$) & 1.88 (2.24) & 0.46 (0.55) \\
%\noalign{\smallskip}
$\dot{E}$ (10$^{-4}$ M$_{\sun}$\,km$^{2}$\,s$^{-2}$ yr$^{-1}$) & 9.61 (13.56) & 1.70 (2.40) \\
L (L$_{\sun}$) & 0.16 (0.22) & 0.03 (0.04) \\
\noalign{\smallskip}
\hline
\end{tabular}    
\label{Tab:Outflow Properites}
\begin{description}
\item The properties given within parentheses have been corrected at an inclination $i$ = 57.3$^{\circ}$, otherwise no correction was applied ($i$ = 0$^{\circ}$).
\end{description}
\end{table}

%__________________________________________________________________

\subsubsection{$^{12}$CO\,(J=1--0) line emission} \label{subsubsec: 12CO results}

\autoref{Fig:CO spectrum} presents the  $^{12}$CO\,(J=1--0) line profile of the large-scale outflow from VLA~2. High-velocity line wings are present in the spectrum. The inner, low velocity channels show evidence for self-absorption and/or missing flux.

\begin{figure}
   \centering
   \includegraphics[width=\hsize]{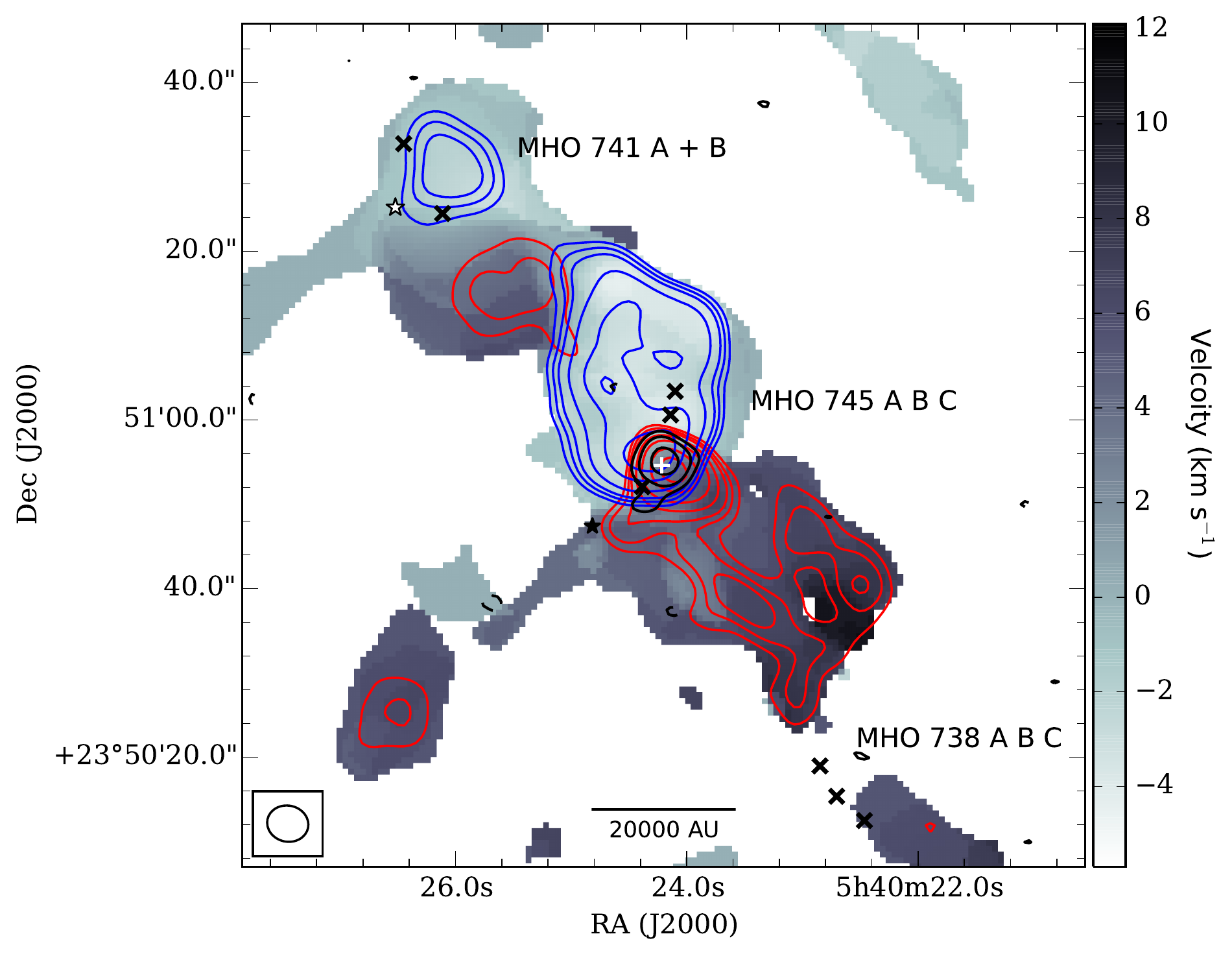}
      \caption{Centred on VLA~2 (marked by a white plus) the contours show $^{12}$CO\,(J=1--0) line emission integrated over -25.9\,-\,1.2~km\,s$^{-1}$ (blue) and 5.3\,-\,20.9~km\,s$^{-1}$ (red) and 2.7-mm continuum emission (black). The contours are over-plotted on a first moment map of the $^{12}$CO\,(J=1--0) emission between -25.9 to 20.9~km\,s$^{-1}$ velocity channels with pixels masked below 1~Jy. Blue-shifted contours are -3, 3, 4, 5, 7, 10, 12, 14 $\times$ 3.0~Jy\,beam$^{-1}$\,km\,s$^{-1}$, red-shifted contours are -3, 3, 4, 5, 7, 10, 15, 20 $\times$ 2.0~Jy\,beam$^{-1}$\,km\,s$^{-1}$ and the black contours are as in \autoref{Fig:mm continuum}. The synthesised beam is shown in the bottom left corner and the scale bar in the bottom middle. MHOs are shown as crosses. A Class I [see \autoref{Fig:3.6 and 6 cm}\,(c and d)] and a Class II object reported by \citet{Gutermuth2009} are shown as a black and white star, respectively.}
      \label{Fig:CO emission}
   \end{figure}

\autoref{Fig:CO emission} shows blue- and red-shifted $^{12}$CO\,(J=1--0) emission integrated over several velocity ranges along with the intensity weighted first moment map of the line emission shown in greyscale. \autoref{Fig:CO emission} also shows six additional markers to that shown in \autoref{Fig:mm continuum}. The three markers about 40~arcsec to the north-east of VLA~2 show the positions of MHOs 741 A and B detected by \citetalias{Khanzadyan2011}, and the white star shows the position of a Class II pre-main-sequence star reported by \citet{Gutermuth2009}. The three remaining markers to the south-west of VLA~2 show the positions of MHOs 738 A, B and C from \citetalias{Khanzadyan2011}, indicated as crosses. 

The $^{12}$CO\,(J=1--0) emission appears to show a bipolar outflow centred on VLA~2, appearing in the same orientation as previously seen by \citet{Zhang2005}. There is also blue- and red-shifted emission $\sim$35~arcsec to the north--east of VLA~2 that appears to resemble another resolved bipolar outflow. However, the only apparent candidate driving source of this outflow is a Class II source reported by \citet{Gutermuth2009}, which does not lie directly between the red and blue lobes of the candidate outflow. Therefore, it is unlikely that this source is driving the outflow and we conclude that the $^{12}$CO\,(J=1--0) emission in fact belongs to the single outflow centred on VLA~2. This could be due to inclination effects. For instance, if the outflow is almost in the plane of the sky, part of the blue-shifted cavity will appear red-shifted.

The groups of MHOs to the north--east and south--west are likely associated with the outflow from VLA~2, as they lie at opposite ends of the outflow seen in $^{12}$CO\,(J=1--0) emission. 

We also detected red-shifted emission extending to the south--east of VLA~2 which ends in a patch of emission at 05$^{\rmn{h}}$40$^{\rmn{m}}$26$\fs$5 +23$\degr$50$\arcmin$52$\farcs$4 (J2000). However, there are no other signposts of star formation at that position.

%____CO line emission (blue/red shifted) &~cm continuum_______

%----------------------------------------------------------- S_vib
%
%______________________________________________________________

\subsubsection{Outflow Properties}

\autoref{Tab:Outflow Properites} presents the derived properties for the red- and blue-shifted lobes of the outflow centred on VLA~2. The rows from top-to-bottom show: velocity range, mass, momentum, kinetic energy, mean velocity, length, dynamical timescale, mass transfer rate, mechanical force and mechanical luminosity for the blue and red lobes of the outflow.

The velocity range shows the range of velocity channels that were integrated over to produce the outflow properties. The inner, low velocity channels were excluded due to self-absorption or missing flux; channels with insignificant emission (< 3$\sigma$) in the high-velocity wings were also excluded.

The outflow mass, $M_{\mathrm{outflow}}$, was derived for each lobe following \citet{Scoville1986}:
\begin{equation}
\centering
M\,(\mathrm{M}_{\sun})= 2.29 \times 10^{-5}\,\frac{(T + 0.926)}{e^{-5.53/T}}\frac{\tau}{\tau - e^{-\tau}}\,d_{\mathrm{kpc}}^2\,\int S_{v}\,dv  
\label{Eqn: Outflow mass}
\end{equation} 
where $\tau$ is the optical depth, which we set to zero, assuming the source to be optically thin, and therefore the third term of \autoref{Eqn: Outflow mass} was set to unity. The distance, $d_{\mathrm{kpc}}$, was assumed to be 1.2~kpc. $S_{v}$ is the integrated flux density in velocity channel, $v$, that was found by summing up the emission within a region defined by the 1$\sigma$\,=\,0.5~Jy\,beam$^{-1}$ contours (the maximal noise level within the spectra). $T$ is the excitation temperature of the outflow gas, which we assumed to be equal to the full Planck function derived brightness temperature, given as: 
\begin{equation}
\centering
T_B\,(\mathrm{K}) = 0.048\,[\nu\,(\mathrm{GHz})] \ln\left(1 + \frac{3.92 \times 10^{-8}\,[\nu\,(\mathrm{GHz})]^{3}\,\theta (\arcsec)^{2}}{F_{\nu}\,(\mathrm{Jy\,beam}^{-1})}\right)^{-1}
\label{Eqn: Brightness Temp}
\end{equation} 
where $\theta$ is the beamsize and $F_{\nu}$ is the peak flux density of the outflow at frequency $\nu$. For the derived properties listed in \autoref{Tab:Outflow Properites} the largest peak flux from either blue- or red-shifted lobe was used to define the brightness temperature for that outflow, which we found to be 6.47~Jy\,beam$^{-1}$ at 5.3~km\,s$^{-1}$ from the red-shifted lobe. We determined the brightness temperature to be 27.3~K and found the total mass contained within the flows to be $\sim$0.8~M$_{\sun}$.

To calculate the outflow momentum, $P$, and kinetic energy, $E$, the $\int$\,$S_{v}$\,d$v$ term in \autoref{Eqn: Outflow mass} was replaced by $\int$\,$S_{v}$\,($v - v_{\mathrm{\,LSR}}$)\,/\,$\cos i$\,d$v$ for the momentum, and $\int$\,$S_{v}$\,[($v - v_{\mathrm{\,LSR}}$)\,/\,$\cos i$\,]$^{2}$\,d$v$ for the kinetic energy, where $v$ is the velocity of each channel, $v_{\mathrm{\,LSR}}$ = 2.3~km\,s$^{-1}$ is the systematic velocity of the cloud taken from \citet{Zhang2005} and $i$ is the inclination of the outflow. 

The columns in \autoref{Tab:Outflow Properites} show the properties uncorrected for inclination ($i$ = 0$\degr$) and the values in parentheses show the properties corrected using a mean inclination angle of $i$ = 57.3$\degr$ \citep{Bontemps1996}. The uncorrected results are more closely relatable to the results of \citet{Zhang2005}, who also assumed an inclination of 0$\degr$. Although we have not applied this, a further correction of 3.5 may be applied to correct for optical depth effects \citep{Bontemps1996}.

The mass weighted velocity, v$_{\mathrm{outflow}}$, of the outflow is given as the $P$/$M_{\mathrm{outflow}}$ of the flow. The length of flow, $L$, was measured from the position of VLA~2 to the furthest 3$\sigma$ contour (see \autoref{Fig:CO emission}) and was corrected for inclination.

From this we define the dynamical timescale, $t_{\mathrm{dyn}}$ = $L$\,/\,v$_{\mathrm{outflow}}$, and compute: the mass transfer rate, $\dot{M}$ = $M_{\mathrm{outflow}}$\,/\,$t_{\mathrm{dyn}}$; the mechanical force, $F$ = $P$\,/\,$t_{\mathrm{dyn}}$; and the mechanical energy transport rate, $\dot{E}$, or mechanical luminosity, L, defined as $E$\,/\,$t_{\mathrm{dyn}}$.

\section{Discussion} \label{Sec:Discussion}

In addition to spectral analysis, a comparison of the radio-to-bolometric luminosity of a source can be used as a means of source classification. For low luminosity objects to have a strong radio flux, the emission can only arise from shock-ionised gas, as there is negligible emission via photoionisation of circumstellar material. At higher luminosities, the expected Lyman continuum flux would dominate the observed radio flux, if the emission were produced via photoionisation i.e. by an H\,\textsc{ii} region. 

%----------------------------------------------------------- S_vib
   \begin{figure}
   \centering
   \includegraphics[trim=0cm 0cm 0cm 0cm, clip=true, width=\hsize]{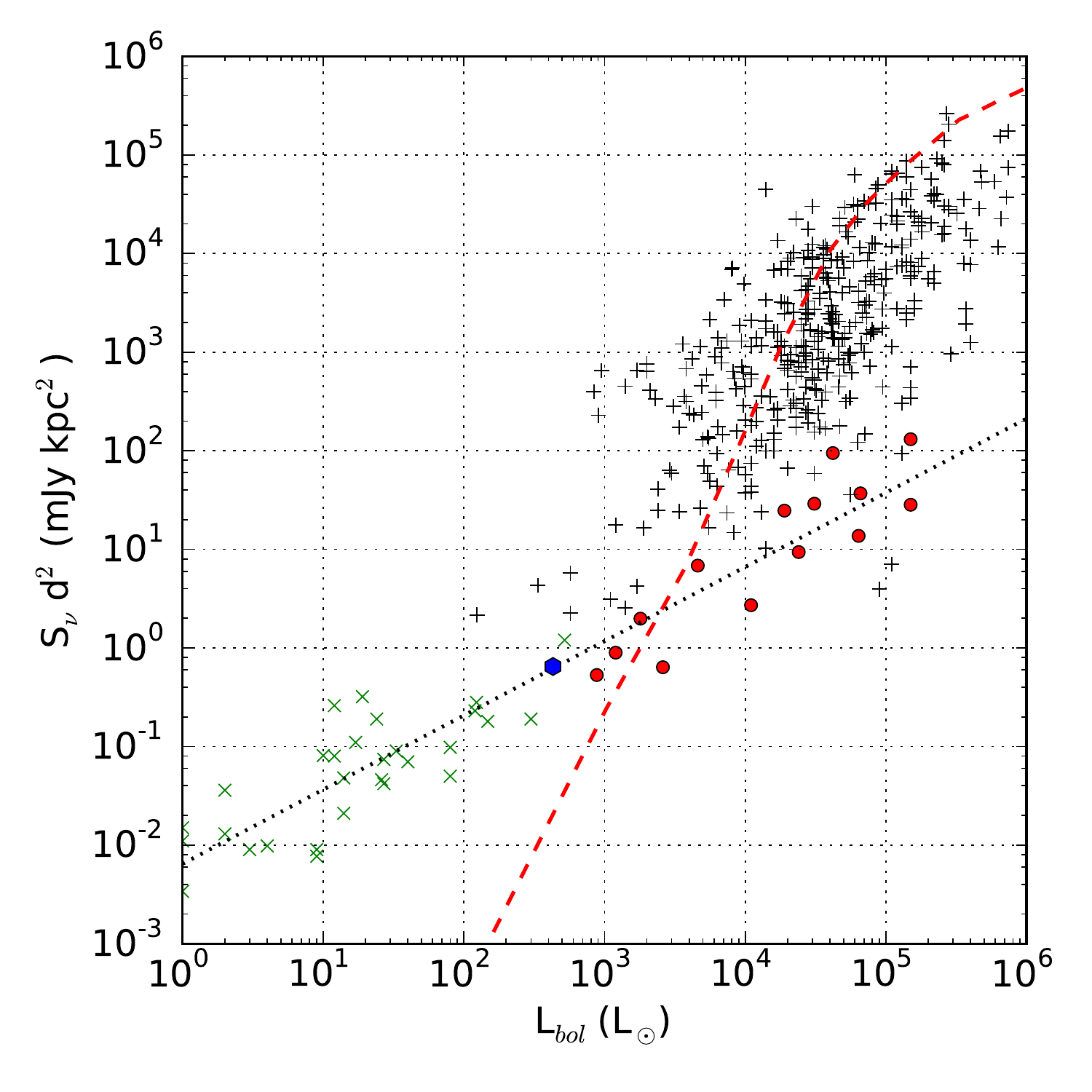}
      \caption{Radio luminosity, $S_\nu$\,$d^2$, at $\nu$ = 5~GHz against the average bolometric luminosity, $L_\mathrm{bol}$, for VLA~2 taken from the literature (indicated as A blue hexagon), H\,\textsc{ii} regions (indicated by black plus signs) and YSO (shown as red circles) detected in the RMS survey \citep{Lumsden2013}. The green crosses indicate low-mass jet-like sources taken from \citet{Anglada1995}. The dashed line shows the results for a single star using stellar models from \citet{LanzHubeny2007}. The dotted line shows the extrapolated jet luminosity fitted to the low-mass stars from \citet{Anglada1995} as well as to known emission from high-mass stars (whether jet or wind) as given in \citet{HoareFranco2007}.}
         \label{Fig:lum_rad vs lum_bol}
   \end{figure}
%______________________________________________________________

\autoref{Fig:lum_rad vs lum_bol} presents the the 6-cm radio flux of VLA~2 ($S_\nu$ = 0.45~mJy) at a bolometric luminosity of 430~L$_{\sun}$, shown as a (blue) hexagon. The bolometric luminosity was found by taking an average of the luminosities found in the literature: 1100~L$_{\sun}$ at a distance of 2.0~kpc \citep[found by the RMS survey,][]{Lumsden2013}, which scales to a luminosity of 396~L$_{\sun}$ at $d$=1.2~kpc; 470~L$_{\sun}$ \citep{Molinari2000}, 400~L$_{\sun}$ \citep{Molinari2008}, 600~L$_{\sun}$ \citep{Zhang2005} and 290~L$_{\sun}$ \citepalias{Khanzadyan2011}; all at a distance of 1.2~kpc.

Also plotted on \autoref{Fig:lum_rad vs lum_bol} are examples of jet-like objects taken from \citet{Anglada1995} shown as green crosses, YSOs and H\,\textsc{ii}-regions detected in the Red MSX Source (RMS) survey \citep{Lumsden2013} shown as red circles and black plus signs, respectively. At the average bolometric luminosity of 430~L$_{\sun}$, VLA~2 appears to be associated with the jet-like objects.

The luminosity of 290~L$_{\sun}$, that was used in our average, was found via SED modelling of VLA~2 \citepalias{Khanzadyan2011}. The authors used a large aperture ($\sim$22000~au) and incorporated fluxes from lower resolution mid-infrared to radio wavelengths in the fitting, which suggested a Class 0/I actively accreting protostar with a mass of about 4.5~M$_{\sun}$. A separate SED fitting yielded a luminosity of 6030~L$_{\sun}$ ($d$ = 1.12~kpc). However, this is likely incorrect as the authors modelled the system as edge on, which greatly increased the fitted luminosity. Thus, this value was not included in our average. 

\subsection{Small-scale Jet} \label{subsec:discussion small Jet}

Multiple shock-features were detected in the region around VLA~2 in H$_2$ line emission \citep[\citetalias{Khanzadyan2011}, also see][]{Varricatt2010}, extending to the north and south, as well as in the north--east south--west directions, roughly aligning with the large-scale outflow we observe in $^{12}$CO\,(J=1--0). 

\begin{figure}
   \centering
   \includegraphics[trim=0cm 0cm 0cm 0cm, clip=true, width=\hsize]{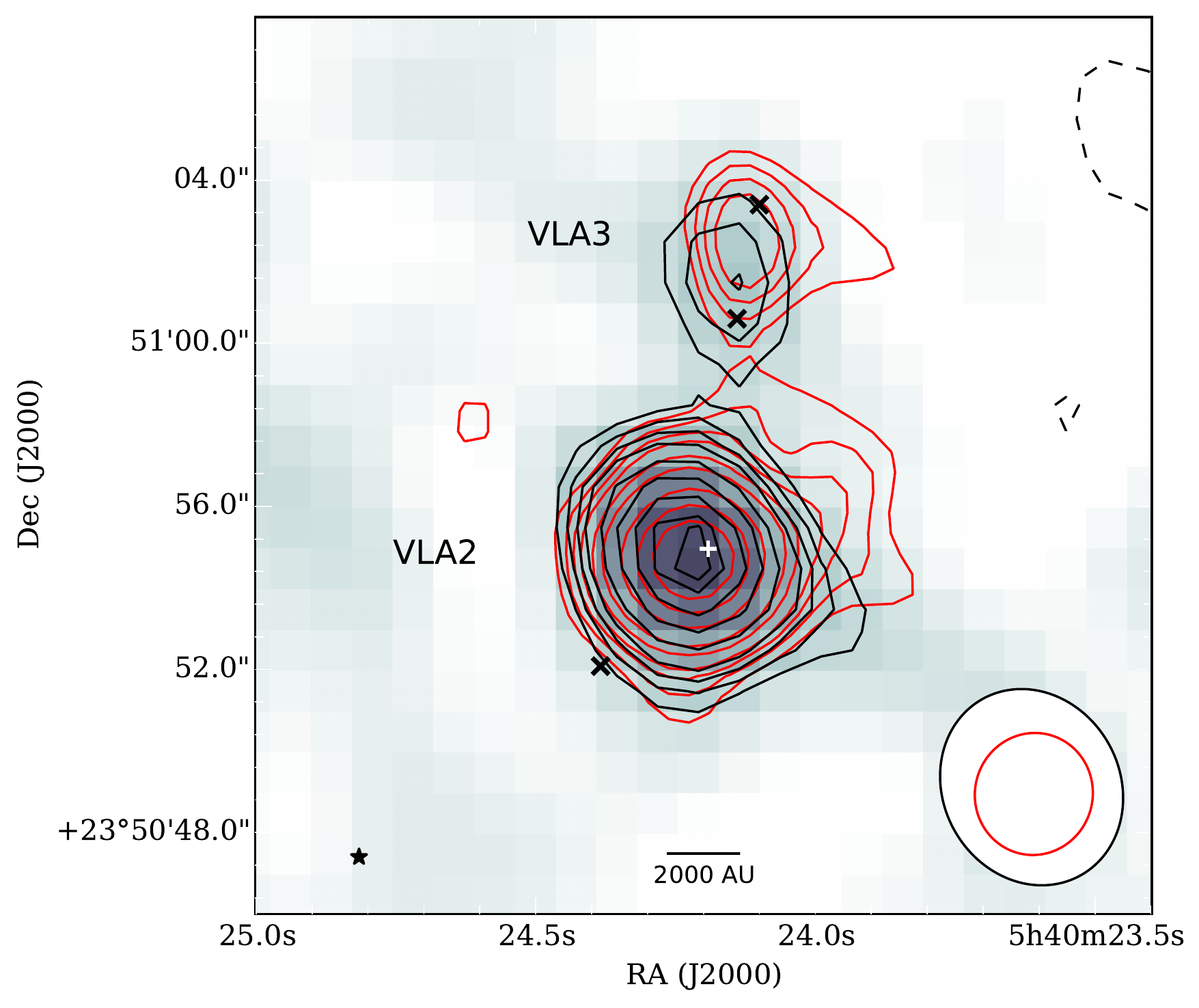}
   
      \caption{Close up of VLA~2 and VLA~3 with 6-cm emission shown by greyscale and black contours at -3, 3, 4, 5 $\times$ 26 $\umu$Jy\,beam$^{-1}$. The red contours show 3.6~cm emission at 3, 4, 5, 7 $\times$ 16 $\umu$Jy\,beam$^{-1}$. The black x's indicate the positions of MHO 745 A, B and C detected by \citet{Khanzadyan2011}. The beams are shown in the bottom left with the larger (black) and smaller (red) ellipse showing the 6 and 3.6-cm beams respectively. The scale bar is also shown in the bottom.}
         \label{Fig:VLA2 MHOs}
   \end{figure}

\autoref{Fig:VLA2 MHOs} presents the positions of MHOs 745 A, B and C, suggested to be powered by a jet centred on VLA~2 \citepalias{Khanzadyan2011}, marked by crosses, along with the contours of the 6 and 3.6-cm emission detected in this work (see \autoref{Fig:3.6 and 6 cm}\,c and d). 

\autoref{Fig:VLA2 MHOs} also shows an offset of $\sim$1~arcsec in the detected peaks of VLA~3 at 6 and 3.6~cm, which equates to $\sim$1200~au at this distance. This suggests a source with varying spectral index or, alternatively, two discrete objects with one being detected only at 6~cm and the other detected only at 3.6~cm. The MHOs also coincide with the blue- and red-shifted $^{12}$CO\,(J=1--0) emission (see \autoref{Fig:CO emission}). This suggests that VLA~2 and the MHOs could also arise from shocks created by jet material impinging upon the cavity walls created by the large-scale outflow.

%------6~cm and 3.6~cm contours overlaid/MHO plot--------- S_vib
%
%______________________________________________________________

\subsection{Large-scale Outflow} \label{subsec:discussion large outflows}

Our $^{12}$CO\,(J=1--0) observations further resolved the previously observed large-scale outflow \citep{Zhang2005} centred on VLA~2. The derived properties for the outflow (listed in \autoref{Tab:Outflow Properites}) are lower than those found by the single dish observations of \citet{Zhang2005}, which were $M_{\mathrm{outflow}}$ = 3.4~M$_{\sun}$, $P$ = 27.1~M$_{\sun}$\,km\,s$^{-1}$ and $E$ = 34 $\times$ 10$^{44}$~ergs. However, this is expected due to the fraction of flux that is recovered by interferometry compared to single dish observations, which we estimate to be around 10 per cent for the line center \citep[e.g.][]{Qiu2009,Kwon2015}.

Interferometric observations of low-mass outflows can expect to find outflow masses on the order of 10$^{-3}$\,--\,10$^{-2}$~M$_{\sun}$ \citep[e.g.][]{Tamura1996}. Single dish observations of massive outflows find outflow masses in the range $\sim$5--1000~M$_{\sun}$ \citep{Maud2015}. Assuming that 10 per cent of the flux in the single dish observations is recovered by the interferometer, we therefore expect massive YSOs to have interferometric outflow masses of 0.5--100~M$_{\sun}$.

Therefore, the outflow detected in $^{12}$CO\,(J=1--0), found to have a total mass of $\sim$0.8~M$_{\sun}$, is higher than that observed for low-mass YSOs and is consistent with the lower end of the expected outflow masses found by \citet{Maud2015}.

\section{Conclusions} \label{Sec:Conclusion}

We present observations of the embedded cluster associated with IRAS~05373+2349 made with the VLA in its C configuration at 6, 3.6 and 1.3-cm wavelengths and CARMA observations of $^{12}$CO(J=1--0) line emission and 2.7-mm continuum. The spectral index of VLA~2 derived from the centimetre observations was found to be consistent with an ionised wind. In addition, the integrated radio flux of VLA~2 compared to the average bolometric luminosity of the associated IRAS source indicated a jet-like object.

We also detected a new radio source, VLA~3, that is coincident with MHOs previously observed in H$_2$ line emission around VLA~2. This supports the claim by \citetalias{Khanzadyan2011} that the MHOs trace a bipolar jet centred on VLA~2, which is roughly aligned with the large-scale outflow presented here in $^{12}$CO(J=1--0). However, we did not find evidence of any of the other outflows in the region, as proposed by \citetalias{Khanzadyan2011}. Therefore, we suggest that the large-scale outflow likely arises from a single jet centred on VLA~2 and traced by VLA~3. Furthermore, the difference in position angle of $\sim$40--50$\degr$ between the jet and the large-scale outflow could be explained by precession of the jet. However, an alternative explanation of VLA~3 could be (shock-)ionisation of the outflow cavity walls of the large-scale outflow.

Finally, our observations of the millimetre continuum and $^{12}$CO\,(J=1--0) emission towards VLA~2 found a core mass of between 12 and 23~M$_{\sun}$ and an outflow mass of 0.8~M$_{\sun}$, intermediate between outflows from low- and high-mass YSOs. 

The goal of this study was to investigate whether the properties of the IM protostar IRAS~05373+2349 VLA~2 are similar to that of low- and/or high-mass protostars. Our observations of the radio continuum emission from IRAS~05373+2349 VLA~2 indicate that it arises in a jet, similar to those observed towards low- and high-mass young stars \citep[e.g.][]{Anglada1995,Purser2016}. In addition, we detect radio continuum emission (VLA~3) associated with previously observed H$_2$ line emission which would indicate shock-ionized knots in the jet from this source, seen commonly in the sample of jets from high-mass YSOs observed by \citet{Purser2016}. The core mass lies between those observed for low- and high-mass YSOs, with the youngest low-mass sources having core masses on the order of a solar mass \citep{Stutz2013}, and the mass reservoirs of high-mass stars reaching thousands of stellar masses \citep{Beuther2002}. The jet itself is accompanied by an outflow from VLA~2, seen in $^{12}$CO (J=1--0) emission, which has properties intermediate between those seen toward low- and high-mass YSOs.

%The above results indicate that the intermediate-mass protostar IRAS~05373+2349 VLA~2 is an example which connects the low- and high-mass regimes of star-formation.

\section*{Acknowledgements}
	We thank David Smith for help with the initial interpretation of the CARMA data. This work was supported in part by the Science and Technology Facilities Council (STFC) grant at the University of Leeds. This work is based in part on observations made with the Spitzer Space Telescope, which is operated by the Jet Propulsion Laboratory, California Institute of Technology under a contract with NASA. This work is based in part on data obtained as part of the UKIRT Infrared Deep Sky Survey.

\bibliography{mybib}

\newcommand{\noop}[1]{}
\begin{thebibliography}{}
\makeatletter
\relax
\def\mn@urlcharsother{\let\do\@makeother \do\$\do\&\do\#\do\^\do\_\do\%\do\~}
\def\mn@doi{\begingroup\mn@urlcharsother \@ifnextchar [ {\mn@doi@}
  {\mn@doi@[]}}
\def\mn@doi@[#1]#2{\def\@tempa{#1}\ifx\@tempa\@empty \href
  {http://dx.doi.org/#2} {doi:#2}\else \href {http://dx.doi.org/#2} {#1}\fi
  \endgroup}
\def\mn@eprint#1#2{\mn@eprint@#1:#2::\@nil}
\def\mn@eprint@arXiv#1{\href {http://arxiv.org/abs/#1} {{\tt arXiv:#1}}}
\def\mn@eprint@dblp#1{\href {http://dblp.uni-trier.de/rec/bibtex/#1.xml}
  {dblp:#1}}
\def\mn@eprint@#1:#2:#3:#4\@nil{\def\@tempa {#1}\def\@tempb {#2}\def\@tempc
  {#3}\ifx \@tempc \@empty \let \@tempc \@tempb \let \@tempb \@tempa \fi \ifx
  \@tempb \@empty \def\@tempb {arXiv}\fi \@ifundefined
  {mn@eprint@\@tempb}{\@tempb:\@tempc}{\expandafter \expandafter \csname
  mn@eprint@\@tempb\endcsname \expandafter{\@tempc}}}

\bibitem[\protect\citeauthoryear{{Anglada}}{{Anglada}}{1995}]{Anglada1995}
{Anglada} G.,  1995, in {Lizano} S.,  {Torrelles} J.~M.,  eds,  Revista
  Mexicana de Astronomia y Astrofisica, vol. 27 Vol. 1, Revista Mexicana de
  Astronomia y Astrofisica Conference Series. p.~67

\bibitem[\protect\citeauthoryear{{Anglada}, {Villuendas}, {Estalella},
  {Beltr{\'a}n}, {Rodr{\'{\i}}guez}, {Torrelles}  \& {Curiel}}{{Anglada}
  et~al.}{1998}]{Anglada1998}
{Anglada} G.,  {Villuendas} E.,  {Estalella} R.,  {Beltr{\'a}n} M.~T.,
  {Rodr{\'{\i}}guez} L.~F.,  {Torrelles} J.~M.,   {Curiel} S.,  1998, \mn@doi
  [\aj] {10.1086/300637}, \href
  {http://adsabs.harvard.edu/abs/1998AJ....116.2953A} {116, 2953}

\bibitem[\protect\citeauthoryear{{Beltr{\'a}n}}{{Beltr{\'a}n}}{2015}]{Beltran2015}
{Beltr{\'a}n} M.~T.,  2015, \mn@doi [\apss] {10.1007/s10509-014-2151-0}, \href
  {http://adsabs.harvard.edu/abs/2015Ap%26SS.355..283B} {355, 283}

\bibitem[\protect\citeauthoryear{{Beltr{\'a}n} \& {de Wit}}{{Beltr{\'a}n} \&
  {de Wit}}{2016}]{Beltran16}
{Beltr{\'a}n} M.~T.,  {de Wit} W.~J.,  2016, \mn@doi [\aapr]
  {10.1007/s00159-015-0089-z}, \href
  {http://adsabs.harvard.edu/abs/2016A%26ARv..24....6B} {24, 6}

\bibitem[\protect\citeauthoryear{{Beltr{\'a}n}, {Girart}, {Estalella}, {Ho}  \&
  {Palau}}{{Beltr{\'a}n} et~al.}{2002}]{Beltran02}
{Beltr{\'a}n} M.~T.,  {Girart} J.~M.,  {Estalella} R.,  {Ho} P.~T.~P.,
  {Palau} A.,  2002, \mn@doi [\apj] {10.1086/340592}, \href
  {http://adsabs.harvard.edu/abs/2002ApJ...573..246B} {573, 246}

\bibitem[\protect\citeauthoryear{{Beltr{\'a}n}, {Girart}  \&
  {Estalella}}{{Beltr{\'a}n} et~al.}{2006}]{Beltran06}
{Beltr{\'a}n} M.~T.,  {Girart} J.~M.,   {Estalella} R.,  2006, \mn@doi [\aap]
  {10.1051/0004-6361:20065132}, \href
  {http://adsabs.harvard.edu/abs/2006A%26A...457..865B} {457, 865}

\bibitem[\protect\citeauthoryear{{Beltr{\'a}n}, {Estalella}, {Girart}, {Ho}  \&
  {Anglada}}{{Beltr{\'a}n} et~al.}{2008}]{Beltran08}
{Beltr{\'a}n} M.~T.,  {Estalella} R.,  {Girart} J.~M.,  {Ho} P.~T.~P.,
  {Anglada} G.,  2008, \mn@doi [\aap] {10.1051/0004-6361:20078045}, \href
  {http://adsabs.harvard.edu/abs/2008A%26A...481...93B} {481, 93}

\bibitem[\protect\citeauthoryear{{Beuther}, {Schilke}, {Menten}, {Motte},
  {Sridharan}  \& {Wyrowski}}{{Beuther} et~al.}{2002}]{Beuther2002}
{Beuther} H.,  {Schilke} P.,  {Menten} K.~M.,  {Motte} F.,  {Sridharan} T.~K.,
   {Wyrowski} F.,  2002, \mn@doi [\apj] {10.1086/338334}, \href
  {http://adsabs.harvard.edu/abs/2002ApJ...566..945B} {566, 945}

\bibitem[\protect\citeauthoryear{{Beuther}, {Churchwell}, {McKee}  \&
  {Tan}}{{Beuther} et~al.}{2007}]{Beuther07}
{Beuther} H.,  {Churchwell} E.~B.,  {McKee} C.~F.,   {Tan} J.~C.,  2007,
  Protostars and Planets V, \href
  {http://adsabs.harvard.edu/abs/2007prpl.conf..165B} {pp 165--180}

\bibitem[\protect\citeauthoryear{{Bontemps}, {Andre}, {Terebey}  \&
  {Cabrit}}{{Bontemps} et~al.}{1996}]{Bontemps1996}
{Bontemps} S.,  {Andre} P.,  {Terebey} S.,   {Cabrit} S.,  1996, \aap, \href
  {http://adsabs.harvard.edu/abs/1996A%26A...311..858B} {311, 858}

\bibitem[\protect\citeauthoryear{{Condon}}{{Condon}}{1984}]{Condon1984}
{Condon} J.~J.,  1984, \mn@doi [\apj] {10.1086/162705}, \href
  {http://adsabs.harvard.edu/abs/1984ApJ...287..461C} {287, 461}

\bibitem[\protect\citeauthoryear{{Crimier} et~al.,}{{Crimier}
  et~al.}{2010}]{Crimier10}
{Crimier} N.,  et~al., 2010, \mn@doi [\aap] {10.1051/0004-6361/200913499},
  \href {http://adsabs.harvard.edu/abs/2010A%26A...516A.102C} {516, A102}

\bibitem[\protect\citeauthoryear{{Fazio} et~al.,}{{Fazio} et~al.}{2004}]{IRAC}
{Fazio} G.~G.,  et~al., 2004, \mn@doi [\apjs] {10.1086/422843}, \href
  {http://adsabs.harvard.edu/abs/2004ApJS..154...10F} {154, 10}

\bibitem[\protect\citeauthoryear{{Fontani}, {Zhang}, {Caselli}  \&
  {Bourke}}{{Fontani} et~al.}{2009}]{Fontani09}
{Fontani} F.,  {Zhang} Q.,  {Caselli} P.,   {Bourke} T.~L.,  2009, \mn@doi
  [\aap] {10.1051/0004-6361/200911617}, \href
  {http://adsabs.harvard.edu/abs/2009A%26A...499..233F} {499, 233}

\bibitem[\protect\citeauthoryear{{Fuente}, {Neri}, {Mart{\'{\i}}n-Pintado},
  {Bachiller}, {Rodr{\'{\i}}guez-Franco}  \& {Palla}}{{Fuente}
  et~al.}{2001}]{Fuente01}
{Fuente} A.,  {Neri} R.,  {Mart{\'{\i}}n-Pintado} J.,  {Bachiller} R.,
  {Rodr{\'{\i}}guez-Franco} A.,   {Palla} F.,  2001, \mn@doi [\aap]
  {10.1051/0004-6361:20000358}, \href
  {http://adsabs.harvard.edu/abs/2001A%26A...366..873F} {366, 873}

\bibitem[\protect\citeauthoryear{{Fuente}, {Ceccarelli}, {Neri}, {Alonso-Albi},
  {Caselli}, {Johnstone}, {van Dishoeck}  \& {Wyrowski}}{{Fuente}
  et~al.}{2007}]{Fuente07}
{Fuente} A.,  {Ceccarelli} C.,  {Neri} R.,  {Alonso-Albi} T.,  {Caselli} P.,
  {Johnstone} D.,  {van Dishoeck} E.~F.,   {Wyrowski} F.,  2007, \mn@doi [\aap]
  {10.1051/0004-6361:20077297}, \href
  {http://adsabs.harvard.edu/abs/2007A%26A...468L..37F} {468, L37}

\bibitem[\protect\citeauthoryear{{Fuente} et~al.,}{{Fuente}
  et~al.}{2009}]{Fuente09}
{Fuente} A.,  et~al., 2009, \mn@doi [\aap] {10.1051/0004-6361/200912623}, \href
  {http://adsabs.harvard.edu/abs/2009A%26A...507.1475F} {507, 1475}

\bibitem[\protect\citeauthoryear{{Gutermuth}, {Megeath}, {Pipher}, {Williams},
  {Allen}, {Myers}  \& {Raines}}{{Gutermuth} et~al.}{2005}]{Gutermuth05}
{Gutermuth} R.~A.,  {Megeath} S.~T.,  {Pipher} J.~L.,  {Williams} J.~P.,
  {Allen} L.~E.,  {Myers} P.~C.,   {Raines} S.~N.,  2005, \mn@doi [\apj]
  {10.1086/432460}, \href {http://adsabs.harvard.edu/abs/2005ApJ...632..397G}
  {632, 397}

\bibitem[\protect\citeauthoryear{{Gutermuth}, {Megeath}, {Myers}, {Allen},
  {Pipher}  \& {Fazio}}{{Gutermuth} et~al.}{2009}]{Gutermuth2009}
{Gutermuth} R.~A.,  {Megeath} S.~T.,  {Myers} P.~C.,  {Allen} L.~E.,  {Pipher}
  J.~L.,   {Fazio} G.~G.,  2009, \mn@doi [\apjs] {10.1088/0067-0049/184/1/18},
  \href {http://adsabs.harvard.edu/abs/2009ApJS..184...18G} {184, 18}

\bibitem[\protect\citeauthoryear{{Hoare} \& {Franco}}{{Hoare} \&
  {Franco}}{2007}]{HoareFranco2007}
{Hoare} M.~G.,  {Franco} J.,  2007, preprint, \href
  {http://adsabs.harvard.edu/abs/2007arXiv0711.4912H} {} (\mn@eprint {arXiv}
  {0711.4912})

\bibitem[\protect\citeauthoryear{{Ikarashi} et~al.,}{{Ikarashi}
  et~al.}{2011}]{Ikarashi2011}
{Ikarashi} S.,  et~al., 2011, \mn@doi [\mnras]
  {10.1111/j.1365-2966.2011.18918.x}, \href
  {http://adsabs.harvard.edu/abs/2011MNRAS.415.3081I} {415, 3081}

\bibitem[\protect\citeauthoryear{Khanzadyan, Movsessian, Davis, Magakian,
  Gredel  \& Nikogossian}{Khanzadyan et~al.}{2011}]{Khanzadyan2011}
Khanzadyan T.,  Movsessian T.~A.,  Davis C.~J.,  Magakian T.~Y.,  Gredel R.,
  Nikogossian E.~H.,  2011, \mn@doi [Monthly Notices of the Royal Astronomical
  Society] {10.1111/j.1365-2966.2011.19618.x}, 418, 1994

\bibitem[\protect\citeauthoryear{{Kwon}, {Fern{\'a}ndez-L{\'o}pez}, {Stephens}
  \& {Looney}}{{Kwon} et~al.}{2015}]{Kwon2015}
{Kwon} W.,  {Fern{\'a}ndez-L{\'o}pez} M.,  {Stephens} I.~W.,   {Looney} L.~W.,
  2015, \mn@doi [\apj] {10.1088/0004-637X/814/1/43}, \href
  {http://adsabs.harvard.edu/abs/2015ApJ...814...43K} {814, 43}

\bibitem[\protect\citeauthoryear{{Lanz} \& {Hubeny}}{{Lanz} \&
  {Hubeny}}{2007}]{LanzHubeny2007}
{Lanz} T.,  {Hubeny} I.,  2007, \mn@doi [\apjs] {10.1086/511270}, \href
  {http://adsabs.harvard.edu/abs/2007ApJS..169...83L} {169, 83}

\bibitem[\protect\citeauthoryear{{Lumsden}, {Hoare}, {Urquhart}, {Oudmaijer},
  {Davies}, {Mottram}, {Cooper}  \& {Moore}}{{Lumsden}
  et~al.}{2013}]{Lumsden2013}
{Lumsden} S.~L.,  {Hoare} M.~G.,  {Urquhart} J.~S.,  {Oudmaijer} R.~D.,
  {Davies} B.,  {Mottram} J.~C.,  {Cooper} H.~D.~B.,   {Moore} T.~J.~T.,  2013,
  \mn@doi [\apjs] {10.1088/0067-0049/208/1/11}, \href
  {http://adsabs.harvard.edu/abs/2013ApJS..208...11L} {208, 11}

\bibitem[\protect\citeauthoryear{{Maud}, {Moore}, {Lumsden}, {Mottram},
  {Urquhart}  \& {Hoare}}{{Maud} et~al.}{2015}]{Maud2015}
{Maud} L.~T.,  {Moore} T.~J.~T.,  {Lumsden} S.~L.,  {Mottram} J.~C.,
  {Urquhart} J.~S.,   {Hoare} M.~G.,  2015, \mn@doi [\mnras]
  {10.1093/mnras/stv1635}, \href
  {http://adsabs.harvard.edu/abs/2015MNRAS.453..645M} {453, 645}

\bibitem[\protect\citeauthoryear{{McMullin}, {Waters}, {Schiebel}, {Young}  \&
  {Golap}}{{McMullin} et~al.}{2007}]{CASAREF}
{McMullin} J.~P.,  {Waters} B.,  {Schiebel} D.,  {Young} W.,   {Golap} K.,
  2007, in {Shaw} R.~A.,  {Hill} F.,   {Bell} D.~J.,  eds,  Astronomical
  Society of the Pacific Conference Series Vol. 376, Astronomical Data Analysis
  Software and Systems XVI. p.~127

\bibitem[\protect\citeauthoryear{{Molinari}, {Brand}, {Cesaroni}  \&
  {Palla}}{{Molinari} et~al.}{2000}]{Molinari2000}
{Molinari} S.,  {Brand} J.,  {Cesaroni} R.,   {Palla} F.,  2000, \aap, \href
  {http://adsabs.harvard.edu/abs/2000A%26A...355..617M} {355, 617}

\bibitem[\protect\citeauthoryear{{Molinari}, {Testi}, {Rodr{\'{\i}}guez}  \&
  {Zhang}}{{Molinari} et~al.}{2002}]{Molinari2002}
{Molinari} S.,  {Testi} L.,  {Rodr{\'{\i}}guez} L.~F.,   {Zhang} Q.,  2002,
  \mn@doi [\apj] {10.1086/339630}, \href
  {http://adsabs.harvard.edu/abs/2002ApJ...570..758M} {570, 758}

\bibitem[\protect\citeauthoryear{{Molinari}, {Pezzuto}, {Cesaroni}, {Brand},
  {Faustini}  \& {Testi}}{{Molinari} et~al.}{2008}]{Molinari2008}
{Molinari} S.,  {Pezzuto} S.,  {Cesaroni} R.,  {Brand} J.,  {Faustini} F.,
  {Testi} L.,  2008, \mn@doi [\aap] {10.1051/0004-6361:20078661}, \href
  {http://adsabs.harvard.edu/abs/2008A%26A...481..345M} {481, 345}

\bibitem[\protect\citeauthoryear{{Neri} et~al.,}{{Neri} et~al.}{2007}]{Neri07}
{Neri} R.,  et~al., 2007, \mn@doi [\aap] {10.1051/0004-6361:20077320}, \href
  {http://adsabs.harvard.edu/abs/2007A%26A...468L..33N} {468, L33}

\bibitem[\protect\citeauthoryear{{Ossenkopf} \& {Henning}}{{Ossenkopf} \&
  {Henning}}{1994}]{Ossenkopf1994}
{Ossenkopf} V.,  {Henning} T.,  1994, VizieR Online Data Catalog, \href
  {http://adsabs.harvard.edu/abs/1994yCat..32910943O} {329, 10943}

\bibitem[\protect\citeauthoryear{{Palau} et~al.,}{{Palau}
  et~al.}{2011}]{Palau11}
{Palau} A.,  et~al., 2011, \mn@doi [\apjl] {10.1088/2041-8205/743/2/L32}, \href
  {http://adsabs.harvard.edu/abs/2011ApJ...743L..32P} {743, L32}

\bibitem[\protect\citeauthoryear{{Panagia} \& {Felli}}{{Panagia} \&
  {Felli}}{1975}]{Panagia1975}
{Panagia} N.,  {Felli} M.,  1975, \aap, \href
  {http://adsabs.harvard.edu/abs/1975A%26A....39....1P} {39, 1}

\bibitem[\protect\citeauthoryear{{Purser}, {Lumsden}, {Hoare}, {Urquhart},
  {Cunningham}, {Garay}  \& {Guzm{\'a}n}}{{Purser} et~al.}{2016}]{Purser2016}
{Purser} S.~J.,  {Lumsden} S.~L.,  {Hoare} M.~G.,  {Urquhart} J.~S.,
  {Cunningham} N.,  {Garay} G.,   {Guzm{\'a}n} A.~E.,  \noop{2016}submitted
  2016, MNRAS

\bibitem[\protect\citeauthoryear{{Qiu}, {Zhang}, {Wu}  \& {Chen}}{{Qiu}
  et~al.}{2009}]{Qiu2009}
{Qiu} K.,  {Zhang} Q.,  {Wu} J.,   {Chen} H.-R.,  2009, \mn@doi [\apj]
  {10.1088/0004-637X/696/1/66}, \href
  {http://adsabs.harvard.edu/abs/2009ApJ...696...66Q} {696, 66}

\bibitem[\protect\citeauthoryear{{S{\'a}nchez-Monge}, {Palau}, {Estalella},
  {Kurtz}, {Zhang}, {Di Francesco}  \& {Shepherd}}{{S{\'a}nchez-Monge}
  et~al.}{2010}]{Sanchez-Monge10}
{S{\'a}nchez-Monge} {\'A}.,  {Palau} A.,  {Estalella} R.,  {Kurtz} S.,  {Zhang}
  Q.,  {Di Francesco} J.,   {Shepherd} D.,  2010, \mn@doi [\apjl]
  {10.1088/2041-8205/721/2/L107}, \href
  {http://adsabs.harvard.edu/abs/2010ApJ...721L.107S} {721, L107}

\bibitem[\protect\citeauthoryear{{Saul}}{{Saul}}{2015}]{Saul15}
{Saul} M.,  2015, \mn@doi [\apj] {10.1088/0004-637X/809/1/86}, \href
  {http://adsabs.harvard.edu/abs/2015ApJ...809...86S} {809, 86}

\bibitem[\protect\citeauthoryear{{Scoville}, {Sargent}, {Sanders}, {Claussen},
  {Masson}, {Lo}  \& {Phillips}}{{Scoville} et~al.}{1986}]{Scoville1986}
{Scoville} N.~Z.,  {Sargent} A.~I.,  {Sanders} D.~B.,  {Claussen} M.~J.,
  {Masson} C.~R.,  {Lo} K.~Y.,   {Phillips} T.~G.,  1986, \mn@doi [\apj]
  {10.1086/164086}, \href {http://adsabs.harvard.edu/abs/1986ApJ...303..416S}
  {303, 416}

\bibitem[\protect\citeauthoryear{{Shepherd} \& {Watson}}{{Shepherd} \&
  {Watson}}{2002}]{Shepherd02}
{Shepherd} D.~S.,  {Watson} A.~M.,  2002, \mn@doi [\apj] {10.1086/338138},
  \href {http://adsabs.harvard.edu/abs/2002ApJ...566..966S} {566, 966}

\bibitem[\protect\citeauthoryear{{Simon}, {Ayres}, {Redfield}  \&
  {Linsky}}{{Simon} et~al.}{2002}]{Simon02}
{Simon} T.,  {Ayres} T.~R.,  {Redfield} S.,   {Linsky} J.~L.,  2002, \mn@doi
  [\apj] {10.1086/342941}, \href
  {http://adsabs.harvard.edu/abs/2002ApJ...579..800S} {579, 800}

\bibitem[\protect\citeauthoryear{{Stutz} et~al.,}{{Stutz}
  et~al.}{2013}]{Stutz2013}
{Stutz} A.~M.,  et~al., 2013, \mn@doi [\apj] {10.1088/0004-637X/767/1/36},
  \href {http://adsabs.harvard.edu/abs/2013ApJ...767...36S} {767, 36}

\bibitem[\protect\citeauthoryear{{Takahashi} \& {Ho}}{{Takahashi} \&
  {Ho}}{2012}]{Takahashi12a}
{Takahashi} S.,  {Ho} P.~T.~P.,  2012, \mn@doi [\apjl]
  {10.1088/2041-8205/745/1/L10}, \href
  {http://adsabs.harvard.edu/abs/2012ApJ...745L..10T} {745, L10}

\bibitem[\protect\citeauthoryear{{Takahashi}, {Saigo}, {Ho}  \&
  {Tomida}}{{Takahashi} et~al.}{2012}]{Takahashi12b}
{Takahashi} S.,  {Saigo} K.,  {Ho} P.~T.~P.,   {Tomida} K.,  2012, \mn@doi
  [\apj] {10.1088/0004-637X/752/1/10}, \href
  {http://adsabs.harvard.edu/abs/2012ApJ...752...10T} {752, 10}

\bibitem[\protect\citeauthoryear{{Tamura}, {Ohashi}, {Hirano}, {Itoh}  \&
  {Moriarty-Schieven}}{{Tamura} et~al.}{1996}]{Tamura1996}
{Tamura} M.,  {Ohashi} N.,  {Hirano} N.,  {Itoh} Y.,   {Moriarty-Schieven}
  G.~H.,  1996, \mn@doi [\aj] {10.1086/118164}, \href
  {http://adsabs.harvard.edu/abs/1996AJ....112.2076T} {112, 2076}

\bibitem[\protect\citeauthoryear{{Testi}, {Palla}  \& {Natta}}{{Testi}
  et~al.}{1999}]{Testi99_new}
{Testi} L.,  {Palla} F.,   {Natta} A.,  1999, \aap, 342, 515

\bibitem[\protect\citeauthoryear{{Thompson}, {Moran}  \& {Swenson}}{{Thompson}
  et~al.}{1986}]{Thompson1986}
{Thompson} A.~R.,  {Moran} J.~M.,   {Swenson} G.~W.,  1986, {Interferometry and
  synthesis in radio astronomy}

\bibitem[\protect\citeauthoryear{{Varricatt}, {Davis}, {Ramsay}  \&
  {Todd}}{{Varricatt} et~al.}{2010}]{Varricatt2010}
{Varricatt} W.~P.,  {Davis} C.~J.,  {Ramsay} S.,   {Todd} S.~P.,  2010, \mn@doi
  [\mnras] {10.1111/j.1365-2966.2010.16356.x}, \href
  {http://adsabs.harvard.edu/abs/2010MNRAS.404..661V} {404, 661}

\bibitem[\protect\citeauthoryear{{Yang}, {Gurvits}, {Frey}  \&
  {Lobanov}}{{Yang} et~al.}{2008}]{Yang2008}
{Yang} J.,  {Gurvits} L.~I.,  {Frey} S.,   {Lobanov} A.~P.,  2008, preprint,
  \href {http://adsabs.harvard.edu/abs/2008arXiv0811.2926Y} {} (\mn@eprint
  {arXiv} {0811.2926})

\bibitem[\protect\citeauthoryear{{Zhang}, {Hunter}, {Brand}, {Sridharan},
  {Cesaroni}, {Molinari}, {Wang}  \& {Kramer}}{{Zhang}
  et~al.}{2005}]{Zhang2005}
{Zhang} Q.,  {Hunter} T.~R.,  {Brand} J.,  {Sridharan} T.~K.,  {Cesaroni} R.,
  {Molinari} S.,  {Wang} J.,   {Kramer} M.,  2005, \mn@doi [\apj]
  {10.1086/429660}, \href {http://adsabs.harvard.edu/abs/2005ApJ...625..864Z}
  {625, 864}

\bibitem[\protect\citeauthoryear{{Zinnecker} \& {Yorke}}{{Zinnecker} \&
  {Yorke}}{2007}]{Zinnecker07}
{Zinnecker} H.,  {Yorke} H.~W.,  2007, \mn@doi [\araa]
  {10.1146/annurev.astro.44.051905.092549}, \href
  {http://adsabs.harvard.edu/abs/2007ARA%26A..45..481Z} {45, 481}

\bibitem[\protect\citeauthoryear{{van Kempen} et~al.,}{{van Kempen}
  et~al.}{2016}]{van-Kempen16}
{van Kempen} T.~A.,  et~al., 2016, \mn@doi [\aap]
  {10.1051/0004-6361/201424725}, \href
  {http://adsabs.harvard.edu/abs/2016A%26A...587A..17V} {587, A17}

\makeatother
\end{thebibliography}

% Don't change these lines
\bsp	% typesetting comment
\label{lastpage}
\end{document}